%% file: main.tex
\documentclass[sigconf]{acmart}

\usepackage{booktabs} 

\usepackage{graphicx}
\usepackage{caption}
\usepackage{caption}
\usepackage{natbib}
\usepackage{amsmath}
\usepackage{soul}
\usepackage{url}
\usepackage{enumerate}
\usepackage[utf8]{inputenc}
\usepackage{amsmath}
\usepackage{mathtools}
\usepackage{algorithm}
\usepackage{algorithmic}
\usepackage{subfigure}   
\usepackage{cancel}
\usepackage{comment}
\usepackage{ctable}
\usepackage{mathtools}
\usepackage{balance}
\usepackage{threeparttable}  
\usepackage{subscript}
\usepackage{booktabs}
\usepackage{multirow}
\usepackage{bm}
\usepackage{enumitem}
\usepackage{amsthm}
\usepackage{fixltx2e}
\usepackage{hyperref}
\usepackage{xcolor,colortbl}
\theoremstyle{definition}   
\newtheorem{Definition}{Definition} 

\newtheorem{Problem}{Problem}
\newtheorem{Example}{Example}
\newcommand{\todo}[1]{\textcolor{red}{ToDo: #1}}
\usepackage{geometry}

\AtBeginDocument{%
	\providecommand\BibTeX{{%
			\normalfont B\kern-0.5em{\scshape i\kern-0.25em b}\kern-0.8em\TeX}}}
\setcopyright{iw3c2w3}
\copyrightyear{2021}
\acmYear{2021}

\acmPrice{}
\acmDOI{10.1145/3442381.3450032}
\acmISBN{978-1-4503-8312-7/21/04}

\acmYear{2021}
\acmBooktitle{Proceedings of the Web Conference 2021 (WWW '21), April 19--23, 2021, Ljubljana, Slovenia} 
\acmConference[WWW '21]{Proceedings of the Web Conference 2021}{April 19--23, 2021}{Ljubljana, Slovenia}

\begin{document}
\title[REST: Relational Event-driven Stock Trend Forecasting]{REST: Relational Event-driven Stock Trend Forecasting}
	
\author{Wentao Xu$^{1,4}$*, Weiqing Liu$^{2}$, Chang Xu$^{2}$, Jiang Bian$^{2}$, Jian Yin$^{3,4}$, Tie-Yan Liu$^{2}$}
\thanks{*Work done while Wentao Xu was an intern at Microsoft Research}
\affiliation{\textsuperscript{1}School of Computer Science and Engineering, Sun Yat-sen University, Guangzhou, China}
\affiliation{\textsuperscript{2}Microsoft Research Asia, Beijing, China}
\affiliation{\textsuperscript{3}School of Artificial Intelligence, Sun Yat-sen University, Zhuhai, China}
\affiliation{\textsuperscript{4}Guangdong Key Laboratory of Big Data Analysis and Processing, Guangzhou, China}
\email{{xuwt6@mail2,issjyin@mail}.sysu.edu.cn}
\email{{weiqing.liu, chanx, jiang.bian, tyliu}@microsoft.com}

	%
	%
	
	
	\input{abstract}
	
	%
	%

	\maketitle
	
	\input{introduction}
	\input{related}
	\input{method}
	\input{experiment}

	\input{conclusion}

	\balance
	\bibliographystyle{ACM-Reference-Format}
	\bibliography{www21}
	
\end{document}

%% file: abstract.tex
\begin{abstract}
Stock trend forecasting, aiming at predicting the stock future trends, is crucial for investors to seek maximized profits from the stock market. Many event-driven methods utilized the events extracted from news, social media, and discussion board to forecast the stock trend in recent years. However, existing event-driven methods have two main shortcomings: 1) overlooking the influence of event information differentiated by the stock-dependent properties; 2) neglecting the effect of event information from other related stocks. In this paper, we propose a relational event-driven stock trend forecasting (REST) framework, which can address the shortcoming of existing methods. To remedy the first shortcoming, we propose to model the stock context and learn the effect of event information on the stocks under different contexts. To address the second shortcoming, we construct a stock graph and design a new propagation layer to propagate the effect of event information from related stocks. The experimental studies on the real-world data demonstrate the efficiency of our REST framework. The results of investment simulation show that our framework can achieve a higher return of investment than baselines.
\end{abstract}

%% file: introduction.tex
\section{Introduction}
\label{sec:intro}
Among various investment channels, the stock market has been demonstrating its significant profit potential in the long run. 
Stock trend forecasting, aiming at forecasting the future price trends of stocks, plays as one of the fundamental techniques and has attracted soaring attention from human wisdom~\cite{sutkatti2019stock}.
According to the Efficient Market Hypothesis~\cite{malkiel1970efficient}, the stock price can truly and instantly reflect the stock value, and the significant price fluctuation indicates the reaction to emerging important stock-related information, usually in the form of events. Motivated by this phenomenon, together with a rising variety of sources to obtain published event information, such as news~\cite{nassirtoussi2015text,ding2016knowledge,hu2018listening,duan2018learning,DBLP:conf/cikm/Liu0SZ18,deng2019knowledge,vargas2017deep}, social media~\cite{yang2015twitter,zhou2016can,xu2018stock,DBLP:conf/cikm/WuZSW18}, and discussion board~\cite{li2014effect,nguyen2015sentiment,zimbra2015stakeholder}, many efforts had been investigating how to mine the valuable patterns from the event information for stock trend forecasting~\cite{li2017web}.

\begin{figure}[t]
	\centering
	\includegraphics[width=0.98\columnwidth]{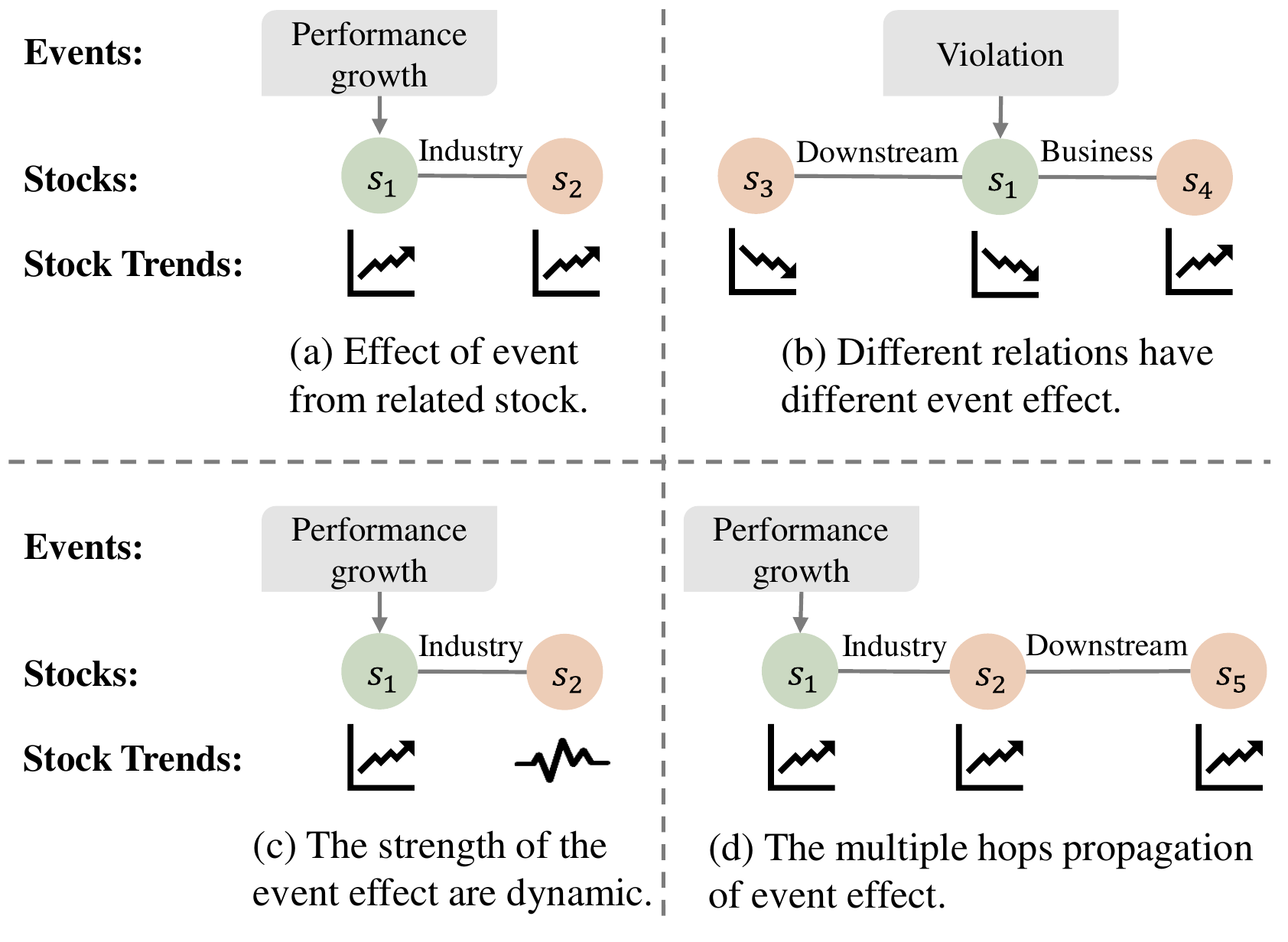}
	\caption{The observations in the propagation of event effect.}
	\label{fig:example}
\end{figure}

However, most of the existing event-driven methods overlook two implicit but critical characteristics of the event information: {\bf Stock-dependent Influence} as well as {\bf Cross-stock Influence}:

\begin{itemize}[leftmargin=1.5em]
	\item {\bf Stock-dependent Influence} 
	 An event's effect on a stock is not only determined by the event but also could be differentiated by the stock-dependent properties. As a result, similar events would have different effects on different stocks. For example, the event of CEO's resignation is undoubtedly a negative signal for a fast-growing company. On the contrary, it is inclined to be interpreted as a positive sign for a company struggling for a long time to find new growth points. However, most existing event-driven methods pay little attention to such crucial stock-dependent event influence. 
	
	\item {\bf Cross-stock Influence}
	Besides directly-associated stocks, an event could yield a similar or even more significant influence on those stocks with no explicit connection.  For example, as shown in Figure~\ref{fig:example}(a), given an event of published performance growth about a company $s_1$, it is a positive signal to the stock price of $s_1$; meanwhile, since this event may indicate a prosperous market environment of the entire industry $s_1$ belonged to, it could exert the similar positive influence to another stock $s_2$ within the same industry. Nonetheless, most of the existing event-driven methods omit such implicit but significant cross-stock influence invoked by events.
\end{itemize} 

To address these shortcomings of existing event-driven methods, in this paper, we propose a novel \underline{r}elational \underline{e}vent-driven \underline{s}tock \underline{t}rend forecasting (REST) framework. Specifically, to consider stock-dependent influence, besides directly modeling the event representations from its detailed textual information, we propose to explore the stock-dependent context of the event for directly related stocks at the same time. This way, we can capture the same event's diverse influence on various stocks given stock-dependent contexts. More concretely, to model the stock context, we take the historical events of this stock into account and consider the subsequent reaction of stock prices to historical events. 
By capturing the fluctuation of one company's stock price and transaction volume after each of its related events, we can model the investors' feedback to this company's historical events, therefore enabling our REST framework to differentiate similar events' influences different companies.

Furthermore, to facilitate modeling the event's cross-stock influence, we construct a stock graph, where the nodes in the graph are stocks, and the edges are the relations between stocks. 
This graph can propagate the effect of event information directly related to one stock to other inter-connected stocks. 
A straightforward way to propagate event information on the stock graph is to leverage the Graph Convolutional Networks (GCN)~\cite{kipf2017semi}. 
However, based on our observations, the propagation of the event effect on the stock graph is complicated, and the traditional GCN can not appropriately model it.
Thus, in our REST framework, we design a new propagation layer to model the complex cross-stock influence inspired by a couple of characteristics observed in the real-world:

\begin{enumerate}[leftmargin=2em]
	\item[{(O1)}] \textbf{An event has variant effects on the stocks connected with different relations.} As shown in Figure~\ref{fig:example}(b), when company $s_1$ was announced that there is a violation in its company operation, its downstream company $s_3$ may bear the same risk of a stock price drop, while its competitor company $s_4$ may contrarily translate this violation exposure event as a positive signal.
	Therefore, our new propagation layer separately models the effect of event propagation through different relations and combines them for the trend prediction.
	
	
	
	\item[{(O2)}] \textbf{The propagation strength of the event effect between two stocks is dynamic.} 
	We observed that the event effect's propagation strength is highly dependent on the dynamically changing contexts of the corresponding pair of related stocks. For example, as shown in Figure~\ref{fig:example}(a), once company $s_1$ announces a remarkable performance growth in the annual report, the public tends to feel optimistic about company $s_2$, in the same growing industry, if the annual report of the company $s_2$ has not released yet.  
	On the other hand, if $s_2$ has already published its annual report before $s_1$, the performance growth of $s_1$ may have very limited influence on $s_2$'s stock price, as shown in Figure~\ref{fig:example}(c). 
	Hence, our new propagation layer models the dynamic propagation strength of influence by taking each stock pair's dynamic context into account.
	
	\item[{(O3)}]\textbf{The effect of the event information could take a multi-hop propagation.} For example, as shown in Figure~\ref{fig:example}(d), an emerging event in terms of brilliant performance growth on stock $s_1$ could imply a positive effect to the stock $s_2$, which is in the same industry as stock $s_1$, as well as another positive signal to stock $s_5$, which is the downstream company of stock $s_2$ and one-more step further to $s_1$. 
	In this paper, we can naturally model the multi-hop influence of events by stacking the propagation layer we designed.
\end{enumerate}


To evaluate our proposed REST framework, we conduct extensive experiments over the real-world data, and experimental results show that our framework can outperform the existing stock trend forecasting methods. Moreover, we simulate the stock investment using a trading strategy, and the investment results show that our framework can achieve a higher investment return than the baselines. 
Finally, we conduct the sensitive tests to further study the effect of different components in our framework and the case study to explain why our framework outperforms the baselines.

In summary, the contributions of our work include:
\begin{itemize}[leftmargin=2em]
	\item We proposed constructing the stock context with the historical events and the corresponding market feed-backs, which is efficient for modeling the various influences of similar events to different stocks.
	
	\item The REST framework we proposed can learn the effect of event information from other related stocks. It can adequately model some essential characteristics we observed in the propagation of event effects.

	\item We conducted both experimental evaluation and investment simulation on real-world data, and the results and analysis verify the validity and rationality of our framework.
\end{itemize}

%% file: related.tex
\section{Related Work}
\label{sec:related}
In recent years, stock trend forecasting has attracted much attention because of its vital stock investment role. We can categorize most of the existing stock trend forecasting work into two categories: the \textbf{Event-driven Methods} and the \textbf{Technical Analysis}. 
\subsection{Event-driven Methods}
According to the efficient market hypothesis~\cite{malkiel1970efficient}, people know that an event that happens on a stock would change the stock information of this stock, affecting its stock price.
There are many efforts to mine the event information from various sources, such as news~\cite{nassirtoussi2015text,ding2016knowledge,hu2018listening,duan2018learning,DBLP:conf/cikm/Liu0SZ18,deng2019knowledge,vargas2017deep}, social media~\cite{si2013exploiting,yang2015twitter,zhou2016can,xu2018stock,DBLP:conf/cikm/WuZSW18}, and discussion board~\cite{li2014effect,nguyen2015sentiment,zimbra2015stakeholder}.
These methods can discover the implicit rules governing the market beyond price data. News is a type of widespread event's source for stock trend forecasting task.~\cite{hu2018listening} proposed to mine news sequence directly from the text with hierarchical attention mechanisms;~\cite{duan2018learning} utilized a target-specific abstract-guided news document representation model to learn the signal in the news thoroughly;~\cite{deng2019knowledge} proposes a novel knowledge-driven temporal convolutional network (KDTCN) to tackle the problem of stock trend prediction and explanation with abrupt changes. Besides, social media (e.g., Twitter) and discussion board (e.g., Yahoo! Finance forum) are also important sources of event information for stock trend forecasting.~\cite{yang2015twitter} proposed a methodology to extract social sentiment from influential Twitter users within a financial community and provided a more robust predictor of financial markets;~\cite{xu2018stock} mined the information in tweets data and used a deep generative model for stock movement prediction;
~\cite{DBLP:conf/cikm/WuZSW18} proposed a novel Cross-model attention based Hybrid Recurrent Neural Network (CH-RNN), which can leverage stock price trend representations to attend daily social text representations through a cross-modal attention interaction;
~\cite{nguyen2015sentiment} predicted stock price movement based on the sentiment analysis of event information in the Yahoo! Finance forum.

Although there were many efforts in exploiting the event information for stock trend forecasting, existing event-driven methods ignore the stock-dependent influence and the cross-stock influence of event information. Our REST framework can overcome the shortages of previous work and learn the stock-dependent influence and cross-stock influence of event information compared with existing methods.
\subsection{Technical Analysis}
Technical analysis~\cite{edwards2018technical} is another category of methods for stock trend forecasting, which is orthogonal to event-driven methods. The technical analysis predicts the stock trend based on the historical time-series of market data, such as trading price and volume. This type of approach aims to discover the trading patterns that can leverage for future predictions. Autoregressive (AR)~\cite{li2016stock} and ARIMA~\cite{ariyo2014stock} models are the most widely used model in this direction, both for linear and stationary time-series. However, the non-linear and non-stationary nature of stock prices limits the applicability of AR and ARIMA models. 
With the recent rapid development of deep learning, some studies attempted to apply deep neural networks to catch the intricate patterns of the market trend~\cite{ticknor2013bayesian,patel2015predicting}. 
To further model the long-term dependency in time series, recurrent neural networks (RNN), especially Long Short-Term Memory (LSTM) network~\cite{hochreiter1997long}, had also been employed in financial perdition~\cite{rather2015recurrent,gao2016stock,akita2016deep,bao2017deep}. Specifically,~\cite{zhang2017stock} proposed a new State Frequency Memory (SFM) recurrent network to discover the multi-frequency trading patterns for stock price movement prediction;~\cite{li2019multi} presented a multi-task recurrent neural network with high-order Markov random fields (MRFs) to predict stock price movement direction;~\cite{feng2019enhancing} leveraged adversarial training to simulate the stochasticity during model training.
Besides, to improve the performance of technical analysis, some recent efforts~\cite{chen2018incorporating,feng2019temporal,kim2019hats,matsunaga2019exploring} leveraged Graph Neural Networks~\cite{kipf2017semi,velickovic2018graph} to capture the relationships between different stocks. 

However, technical analysis is not sensitive to the abrupt changes in stock price caused by external event information of stock~\cite{deng2019knowledge}, limiting its performance on the stock trend forecasting.

%% file: method.tex
\section{Preliminaries}
\label{sec:definition}
In this section, we will introduce some concepts in our proposed relational event-driven stock trend forecasting framework and formally define the problem of stock trend forecasting. 
\begin{Definition}
	\textbf{Stock Context.} The stock market is dynamic, and the stock context of a stock is also dynamic. To better represent the stock context, we define the stock context as the combination of stock's historical events and these events' feedback.
\end{Definition}
\begin{Definition}
	\textbf{Event's Feedback.}
	An event's feedback is the relative change of price and transaction volume on the stock that this event happened.
\end{Definition}

\begin{Example}
Table~\ref{tab:example_feedback} shows the example of the feedback of a performance growth event on the stock \textit{Changan Automobile}, and this event happened on Jan. 28th, 2016. There are $6$ stock price and transaction volume data on a day, which are \textit{opening price} (Open), \textit{closing price} (Close), \textit{highest price} (High), \textit{lowest price} (Low), \textit{trading volume} (Volume) and \textit{volume weighted average price} (VWAP)~\cite{berkowitz1988total}. Thus, this event's feedback on the stock \textit{Changan Automobile} is a $6$-dimensional vector, which is the relative change of price and transaction volume data. The event's feedback can differentiate the influences of similar events on different companies, which is significant for us to learn the distinct effect of event information for each stock.
\end{Example}
\begin{Definition}
	\textbf{Stock Graph.}
	We define the stock graph as a directed graph $G = \left<\mathcal{S}, \mathcal{R}, \mathcal{A} \right>$, where $\mathcal{S}$ denote the set of stocks in the market and $\mathcal{R}$ is the set of relations between two stocks. $\mathcal{A}$ is the set of adjacent matrices. For an  adjacent matrix $A^r \in \mathcal{A}$ ($r \in \mathcal{R}$ and $A^r \in \mathbb{R}^{|\mathcal{S}| \times |\mathcal{S}|}$) of relation $r$, $A^r_{ij} =1$ means there is a relation $r$ from stock $s_j$ to stock $s_i$ and $A^r_{ij} =0$ indicates there is no a relation $r$ from stock $s_j$ to stock $s_i$.
\end{Definition}

\begin{Definition}
	\textbf{Stock Price Trend.} 
	The stock price trend is usually defined as the future change rate of the stock price~\cite{hu2018listening,qin2019you,yang2020html}. In this paper, we define the stock price trend for stock $s_i$ at date $t$ as the stock price change rate of the next day:
	\begin{align}
	d_i^t= \dfrac{Price_i^{t+1} - Price_i^t}{Price_i^t},
	\label{eq:stock-trend}
	\end{align}
\end{Definition}

\noindent where $Price_i^t$ could be specified by different values, such as \textit{opening price}, \textit{closing price} and \textit{volume weighted average price} (VWAP), and we use \textit{closing price} in our work.

\begin{table}[t]
	\centering
	\caption{An example the event's feedback.}
	\resizebox{0.48\textwidth}{!}{
	\begin{tabular}{ p{3.4em}  | c  c   c c c c}
		\hline
		& Open  & Close& High& Low &Volume &VWAP\\
		\hline
		$20160128$ &  $12.34$  & $12.46$& $12.93$& $12.26$ &$364400$ &$12.61$\\
		$20160129$ &  $12.49$  & $13.01$& $13.13$& $12.32$ &$297200$ &$12.88$\\
		\hline
		Feedback &  $1.22\%$ & $4.41\%$& $1.55\%$& $0.49\%$ & $-18.44\%$ &$2.14\%$\\
		\hline
	\end{tabular}}
	\label{tab:example_feedback}
\end{table}

\begin{Problem}
	\textbf{Stock Trend Forecasting.}
	Given the stock-specific information (e.g., the textual information from news and social media, the historical stock price) of stock $s_i$ at date $t$, the goal of stock trend forecasting is to forecast the stock price trend $d_i^t$.
\end{Problem}

\begin{figure*}[t]
	\centering
	\includegraphics[width=2.1\columnwidth]{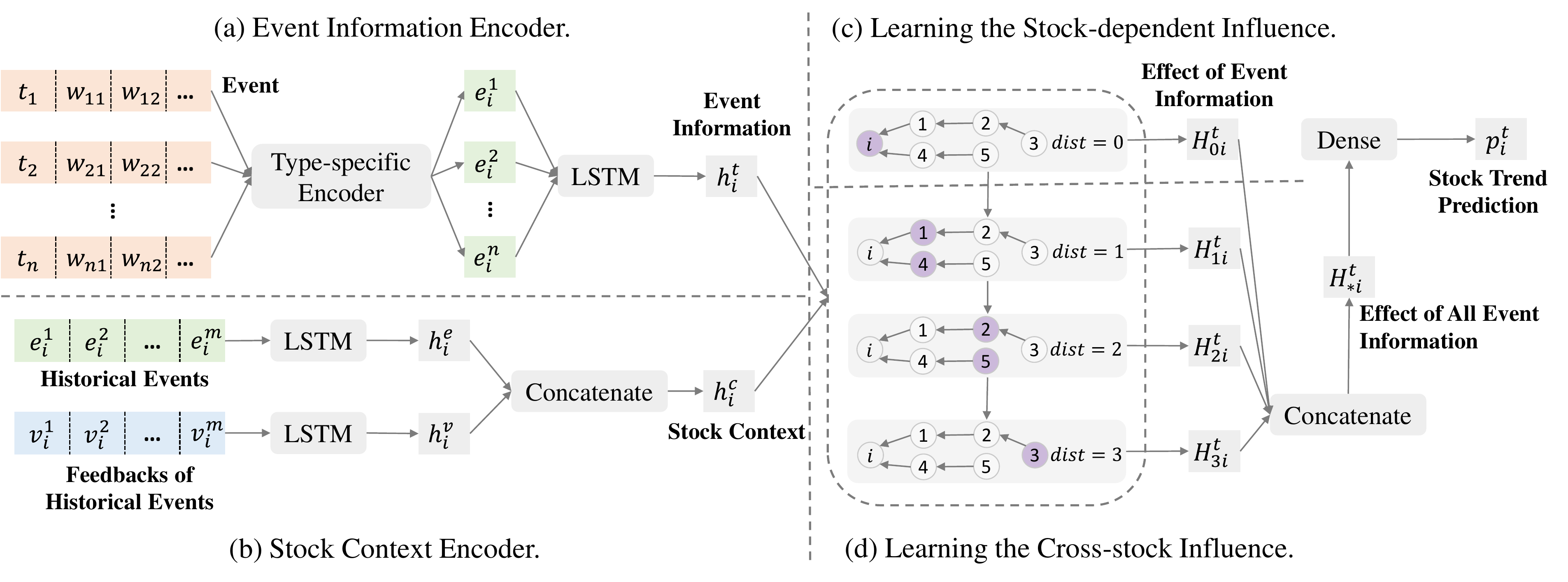}
	\caption{The illustration of the relational event-driven stock trend forecasting framework.}
	\label{fig:framework}
\end{figure*} 

\section{Our REST Framework}
\label{sec:method}
In this section, we elaborate on our relational event-driven stock trend forecasting (REST) framework. Figure~\ref{fig:framework} is the illustration of our REST framework. In Section~\ref{subsec:learn_event}, we utilize an event information encoder to learn the representation of event information. To model the stock-dependent influence, in Section~\ref{subsec: learn_state}, we learn the context of each stock; in Section~\ref{subsec: learn_effect}, we use the event information and the stock context to learn the stock-dependent effect of event information on a stock. After that, to model the cross-stock influence, in Section~\ref{subsec: propagationc}, we propagate the effect of event information on the stock graph with a propagation layer we designed. Our proposed propagation layer can model the three observations in propagating the event effect, which is mentioned in Section~\ref{sec:intro}. 


\subsection{Event Information Encoder}
\label{subsec:learn_event}

We first learn the representation of event information of a stock on a day. Figure~\ref{fig:framework}(a) shows the event information encoder that we use to learn the representation of event information, including the type-specific event encoder and event sequence encoder.

\subsubsection{Type-specific Event Encoder}
\label{sec:encoder}
We utilize a type-specific event encoder to learn the embedding of an event. The events of stocks can categorize into many types according to their content, which may describe the company's performance, the result of litigation, and clarification of rumors. For each event, we have a type embedding $t_i$ and a sequence of token embeddings [$w_{i1}$, $w_{i2}$, ...], where $t_i$ represents the type of this event, and $w_{ix}$ represents the $x$-th token in its content. 
We need to learn which word contributes most to the stock trend for each type of event. We introduce a type-specific multi-head attention mechanism to aggregate the tokens weighted by an assigned attention value, to reward the tokens offering important signal. Specifically,
\begin{align}
	u^k_{ix} =\mathrm{LeakyReLU}\left(W_e^k w_{ix} + b_e^k\right),
\end{align}
\begin{align}
	\alpha^k_{ix} = \dfrac{\exp \left(t_i^\mathsf{T}u^k_{ix}\right)}{\sum_{x=1} \exp \left(t_i^\mathsf{T}u^k_{ix}\right)},
	\label{equ3}
\end{align}
\begin{align}
	e_i =\mathrm{Concat}\left(\sum_x \alpha^1_{ix} w_{ix}, \sum_x \alpha^2_{ix} w_{ix}, ..., \sum_x \alpha^K_{ix} w_{ix}\right).
\end{align} 
We first feed the token $w_{ix}$ through a fully connected layer to get a hidden representation $u^k_{ix}$. We measure the importance of token on this type as the similarity of $u^k_{ix}$ with the type embedding $t_i$ and calculate a normalized attention weight $\alpha^k_{ix}$ through a softmax function. Thus our model can learn the token importance of a specific type of event. Furthermore, we employ \textit{multi-head attention}~\cite{vaswani2017attention} to jointly attend to information from different representation subspaces at different positions. That is, $K$ times independent attention mechanisms execute the calculation of Equation~\ref{equ3} to get the weight $\alpha^k_{ix}$, and then concatenate the weighted sum vector $\sum_x \alpha^k_{ix} w_{ix}$ of tokens to get the event embedding $e_i$.

\subsubsection{Event Sequence Encoder}
Due to the influence of an event on the stock price would last for several days, we should also consider the events in the past couple of days when we predict the stock trend $d_i^t$ of date $t$. For the event embedding sequence [$e_i^1$, $e_i^2$, ..., $e_i^n$] of stock $s_i$ in the latest $3$ days, we feed them into a one-layer LSTM~\cite{hochreiter1997long} and take the last hidden state $h_i^t$ as the representation of event information of stock $s_i$:
\begin{align}
	h_i^t= \mathrm{LSTM}\left(e_i^1, e_i^2, ..., e_i^n\right).
\end{align}
Since the LSTM network can capture the long-term dependency in the events' sequence, the representation $h_i^t$ contains the information of any one of the events in the latest  $3$ days. We can also rewrite the event information of all stocks on date $t$ as a matrix $\bar{H}^t$, where the $i$-th row of $\bar{H}^t$ is the event information $h_i^t$ of stock $s_i$.

\subsection{Stock Context Encoder}
\label{subsec: learn_state}
The stock market is dynamic, and the stock context would influence not only the event information's effect on the directly-associated stock but also the propagation of event effects on the stock graph.
In order to better represent the stock context, we not only use historical events [$e_i^1$, $e_i^2$, ..., $e_i^m$] of stock $s_i$ in the past $30$ days, but also utilize the corresponding feedbacks of historical events [$v_i^1$, $v_i^2$, ..., $v_i^m$]. 
We use the same type-specific event encoder in Section~\ref{sec:encoder} to learn the representation of historical events. 
Moreover, Section~\ref{sec:definition} has introduced the definition of an event's feedback, which is the relative change of stock price and transaction volume after an event happens. 

As shown in Figure~\ref{fig:framework}(b), we feed the sequences of historical events [$e_i^1$, $e_i^2$, ..., $e_i^m$] and feedbacks of historical events [$v_i^1$, $v_i^2$, ..., $v_i^m$] to two different LSTMs, and concatenate the last hidden states $h_i^e$ and $h_i^v$ of these two LSTMs as the stock context $h_i^c$:
\begin{equation}
	\begin{aligned}
		h_i^e &= \mathrm{LSTM}\left(e_i^1, e_i^2, ..., e_i^m\right),\\
		h_i^v &= \mathrm{LSTM}\left(v_i^1, v_i^2, ..., v_i^m\right),\\
		h_i^c &= \mathrm{Concat}\left(h_i^e, h_i^v\right).
	\end{aligned}
\end{equation}
It is crucial to notice that when we forecast the stock trend $d_i^t$ on date $t$, we do not know the stock future price and transaction volume at date $t+1$, so the feedbacks of events happened on the date $t$ is unknown. Therefore, the historical events and the corresponding feedbacks we use to learn the context $h_i^c$ do not include the events that happened on the date $t$. We will explore the influence of the stock context in Section~\ref{sec:effect_state}.

\subsection{Learning the Stock-dependent Influence}
\label{subsec: learn_effect}
To learn the effect of event information on the stocks differentiated by the stock-dependent properties (stock-dependent influence), we utilize the representation of event information and the stock context to learn the effect of one stock's event information on this stock. 
In other words, we learn the strength of event information's effect on a stock.
Formally, we learn the strength of event information's effect from the stock context $h_i^c$ and the event information $h_i^t$:
\begin{align}
D_{ii}^t = \mathrm{LeakyReLU}\left(a^{\mathsf{T}} \left(h_i^c  || h_i^t\right)\right),
\end{align}
where $a^\mathsf{T}$ is a single-layer feed-forward neural network and $||$ represents concatenation. We use the concatetion because it can retain complete information of the event and stock context. $D^t$ is a diagonal matrix where $D_{ii}^t$ is the effect's strength of stock $s_i$'s event information to stock $s_i$. After that, we left multiply event information $\bar{H^t}$ by the strength of event effect $D^t$ to learn the effect of event information:
\begin{align}
 H_0^t = D^t \bar{H^t},
\end{align}
where $H_0^t$ is the effect of event information and its $i$-th row $H_{0i}^t$ is the effect of stock $s_i$'s event information on stock $s_i$.

\subsection{Learning the Cross-stock Influence}
\label{subsec: propagationc}
To address the limitation of existing event-driven work that can not learn the effect of events from other related stocks (cross-stock influence). 
We construct a stock graph defined in Section~\ref{sec:definition}, where the nodes in the graph are stocks, and the edges are stock relations, such as the industry relation and the downstream relation. 
We will introduce the details of the stock relations we used in Section~\ref{subsec:data}.
After that, we design a novel propagation layer to propagate the effect of event information from related stocks in the stock graph. When we design the propagation layer, we pay close attention to modeling the following observations in the propagation of event effect: (O1) different relations result in different event effects; (O2) the dynamic strength of the event effect between two stocks; (O3) the propagation of event effect on multi-hop path. 
Given a matrix $H^t_{0}$ of the effect of event information on date $t$, we progressively design our propagation layer to propagate the effect of event information $H^t_{0}$ and realize to model these three observations.

\subsubsection{GCN Propagation Layer}
\label{subsubsec: GCN}
We first utilize the aggregator in Graph Convolutional Networks (GCN)~\cite{kipf2017semi} to propagate the effect of event information from related stocks:
\begin{align}
H^t_{1}  =\tilde{A}H^t_{0} = D^{-\frac{1}{2}} A D^{-\frac{1}{2}}H^t_{0},
\end{align}
where $\tilde{A} = D^{-\frac{1}{2}} A D^{-\frac{1}{2}}$ and $A$ is the adjacent matrix for all relations, that is, $A_{ij} =1$ indicates there is an arbitrary relation from stock $s_j$ to stock $s_i$. $D$ is a diagonal degree matrix with entries $D_{ii} = \sum_j A_{ij}$, and $D^{-\frac{1}{2}}$ is used to normalize $A$.

After the GCN propagation, the $i$-th row of $H^t_{1}$ contains all the effect of event information of stock $s_i$'s neighbor stocks. Then we can learn the effect of events of related stocks from the matrix $H^t_{1}$.

\subsubsection{Relational GCN Propagation Layer (O1)}
\label{subsubsec: RGCN}
Different relations between two stocks result in different event effects, but the straightforward GCN propagation can not distinguish the impacts of different relations. To remedy this problem, we adopt the idea in the relational GCN~\cite{schlichtkrull2018modeling} to learn the influence of different relations:
\begin{align}
H^t_{1}  =\sum_{r \in \mathcal{R}} \tilde{A}^rW_r H^t_{0}  = \sum_{r \in \mathcal{R}} D^{-\frac{1}{2}} A^r D^{-\frac{1}{2}} W_r H^t_{0},
\end{align}
where $A^r$ is the adjacent matrix of relation $r$, and $W_r$ is the mapping matrix for relation $r$. In the relational GCN propagation layer, we map the effect of event information $H^t_{0}$ with a relation-specific mapping matrix $W_r$ before propagating the event effect, and $W_r H^t_{0}$ represents the event effect propagated with relation $r$. Thus, the relational GCN propagation layer can learn the propagation of event effects under different relations.

\subsubsection{Dynamic Propagation Layer (O2)}
\label{subsubsec: Dynamic_Propagation}
The GCN propagation layer and relational GCN propagation layer use a fixed weight $\tilde{A}_{ij}$ or $\tilde{A}^r_{ij}$ to propagate the event effect from stock $s_j$ to stock $s_i$. 
However, the real-world stock market is dynamic, and the strength of the event effect between two stocks is also dynamic.
To model this observation, we propose to use the dynamic propagated weights to propagate the effect of event information:
\begin{align}
H^t_{1}  = \sum_{r \in \mathcal{R}} \bar{A}^rW_r H^t_{0},
\end{align}
where $\bar{A}^r$ is the matrix of dynamic propagated weights.

We utilize the stock context to learn the dynamic weights $\bar{A}^r$. 
We learn the dynamic propagated weight $\bar{A}^r_{ij}$ of relation $r$ from the stock $s_i$'s neighbor stock $s_j$ ($A^r_{ij}=1$)  to $s_i$ with the context of these two stocks:
\begin{align}
\bar{A}^r_{ij} = f\left( h_i^c, h_j^c\right) = \mathrm{LeakyReLU}\left(b_r^{\mathsf{T}} \left( h_i^c || h_j^c\right)\right).
\end{align}
The function $f(\cdot)$ can have many different forms, in this paper, we adopt the form of function $f(\cdot)$  shown above, where $b_r^\mathsf{T}$ is a single-layer feed-forward neural network for the relation $r$ and $||$ is the concatenation. If a stock $s_k$ is not the neighbor of stock $s_i$ ($A^r_{ik}=0$), we set the propagated weight $\bar{A}^r_{ik}$ as $0$.

\subsubsection{Multi-hop Propagation (O3)}
\label{subsubsec:multi-hop}
The effect of event information will propagate through the multi-hop path. Therefore, we stack more propagation layers to gather the neighboring stocks' event effect with a distance of more than one hop. More formally, we propagate the event effect $H^t_{l-1}$ from the neighbor stocks of $l-1$ hops to gather the event effect from the neighbor stocks of $l$ hops:
\begin{align}
H^t_{l}  = \sum_{r \in \mathcal{R}} \bar{A}^r W_r H^t_{l-1}.
\end{align}

As shown in Figure~\ref{fig:framework} (d), after performing $l$ propagation layers, we obtain the event effect from the neighbor stocks of  different distances: [$H^t_{1}$, $H^t_2$, ..., $H_l^t$]. To jointly learn the effect of original event information $H^t_{0}$ and the event effect of the neighbors of different distances, we leverage the layer-aggregation mechanism~\cite{pmlr-v80-xu18c} to concatenate the effect of all event information of each stock:
\begin{align}
H^t_*  = \mathrm{Concat}\left(H^t_{0}, H^t_{1}, ..., H^t_{l}\right).
\end{align}
By doing so, we cannot only learn the event effect from the neighbors of different distances but also can control the distance of event effect propagation by adjusting the hyper-parameter $l$. We will study the effect of the propagating distance $l$ in Section~\ref{sec:effect_dist}.

\subsection{Stock Trend Forecasting}
We feed the effect of all event information $H^t_{*i}$ (the $i$-th row of the matrix $H^t_*$) of stock $s_i$ to a dense layer, and output our REST framework's stock trend forecasting $p^t_i$  of stock $s_i$ on date $t$:
\begin{align}
p^t_i = W H^t_{*i} + b.
\end{align}

\subsection{Optimization}
We leverage the stochastic gradient descent (SGD) algorithm to optimize our REST framework by minimizing the mean squared error (MSE) loss function with $L_2$ regularization:
\begin{align}
  \mathcal{L} =\frac{1}{|\mathcal{T}|} \sum_{t \in \mathcal{T}}  \sum_{s_i \in \mathcal{S}^{t}} \frac{\left( p_i^t - d_i^t \right)^2 }{|\mathcal{S}^{t} |} + \lambda ||\Theta||^2_2,
\end{align}
where $\mathcal{T}$ is the set of dates in the training period, and $\mathcal{S}^{t}$ is the set of stocks in date $t$. The $p_i^t$ is the stock trend prediction of stock $s_i$ at date $t$, and the $d_i^t$ is the ground truth stock trend of stock $s_i$ at date $t$. $\lambda$ is the regularization parameter; $\Theta$ represents all of the parameters in our framework, including the parameters in neural networks, type embeddings, and token embeddings. 

%% file: experiment.tex
\section{EXPERIMENTS}
\label{sec:experiment}
In this section, we study our relational event-driven stock trend forecasting framework with experiments. We aim to answer the following research questions via the experiments:
\begin{itemize}[leftmargin=1.5em]
	\item \textbf{RQ1}: How is the performance of our REST framework?
	\item \textbf{RQ2}: Can our framework achieve a higher investment return in the investment simulation on real-world datasets?
	\item \textbf{RQ3}: How is the effect of different components (i.e., the stock context, the multi-hop propagation) in our framework? 
	\item \textbf{RQ4}: How does the REST perform in real-world case?
\end{itemize}

\subsection{Data Collection and Preprocessing}
\label{subsec:data}
\paragraph{Datasets}
We evaluate our REST framework on the stocks in two popular and representative stock indexes: CSI 300 and CSI 500. CSI 300 index consists of the $300$ largest and most liquid stocks, reflecting the market's overall performance. CSI 500 index consists of the largest remaining $500$ stocks after excluding the CSI 300 Index constituents, reflecting the small-mid cap stocks.
We collected the stock data, including the event data, stock relations, event's feedback, and the label of stock price trend from 2013 to 2018. We split the stock data by time to a training period from 2013 to 2016, a validation period of 2017, and a test period of 2018. 

\paragraph{Event Data}
We collected the event information from the company's announcements\footnote{The announcement data are crawled from a company data website: \url{http://www.cninfo.com.cn}.}, which have predefined types and have less noise than the event information extracted in the news or social media. We extracted its company name, type of content, content (abstract), and publication timestamp as an event for each announcement. We clear the content of announcements by converting decimals to integers and removing the tokens whose frequency is lower than 5; therefore, we can vastly reduce the scale of token vocabulary. The event data we collected are from 2013 to 2018, and we collected $100629$ and $183325$ events for the stocks in the CSI 300 index and CSI 500 index, respectively.

\paragraph{Stock Relations}
To construct the stock graphs for the stocks in CSI 300 and CSI 500 indexes, we collect four types of stock relations\footnote{We collect the stock relations from a publicly available API tushare: \url{https://tushare.pro/}.}, which are listed as follows:
\begin{itemize}[label={-}, leftmargin=1.5em]
	\item Industry: there is an industry relation between two stocks when they are in the same industry (e.g., Car, Bank, and Electronics).
	\item Business: there is a business relation between two stocks when they have the same business. (e.g., wine, phone, and stainless steel)
	\item Shareholder: we only consider the top $10$ shareholders of each stock. There is a shareholder relation between two stocks when these two stocks have the same shareholder.
	\item Downstream/Upstream: there is a downstream/upstream relation between two stocks when these two stocks' business has a downstream/upstream relation.
\end{itemize}

\paragraph{Event's Feedback} 
To obtain each event's feedback, we collect the daily stock price and volume data of stocks in CSI 300 and CSI 500 indexes\footnote{We collect daily stock price and volume data from \url{http://xueqiu.com}.}. There are 6 stock price and volume data on each day, which are the \textit{opening price}, \textit{closing price}, \textit{highest price}, \textit{lowest price}, \textit{volume weighted average price} (VWAP) and \textit{trading volume}. We calculate the relative change of stock price and volume data after an event happened as the event's feedback. Thus, the feedback of each event is a 6-dimensional vector. 

\paragraph{Label of the Stock Price Trend}
We use the stock price trend defined on Equation~\ref{eq:stock-trend} as the stock trend label on each day. For better training our framework, we apply normalization on the original labels of the same date.
Specifically, we calculate the mean and standard deviation of stock labels on the same date. Then, we subtract the mean and divide by the standard deviation for each stock's label on this date.
\begin{figure*}[t]
	\includegraphics[width=1.9\columnwidth]{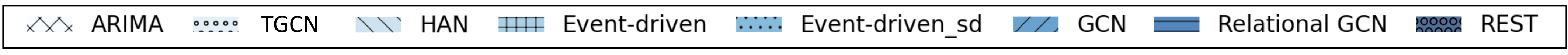}
	\vspace{-.1in}
\end{figure*} 
\begin{figure*}[t]
	\centering
	\subfigure[The Annual Return on CSI 300.]{\includegraphics[width=0.51\columnwidth]{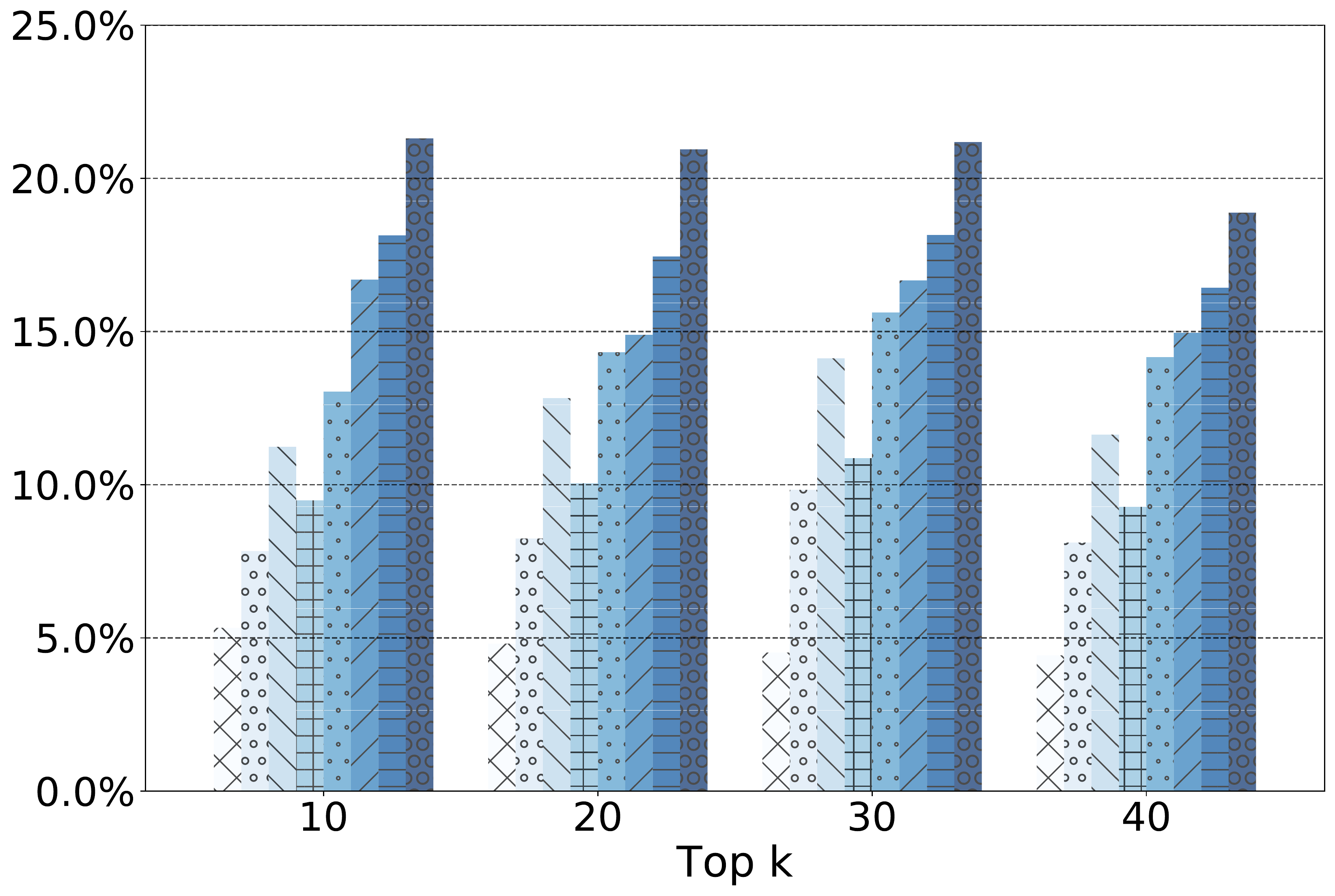}}
	\hspace{.05in} 
	\subfigure[The Annual Return on CSI 500.]{\includegraphics[width=0.51\columnwidth]{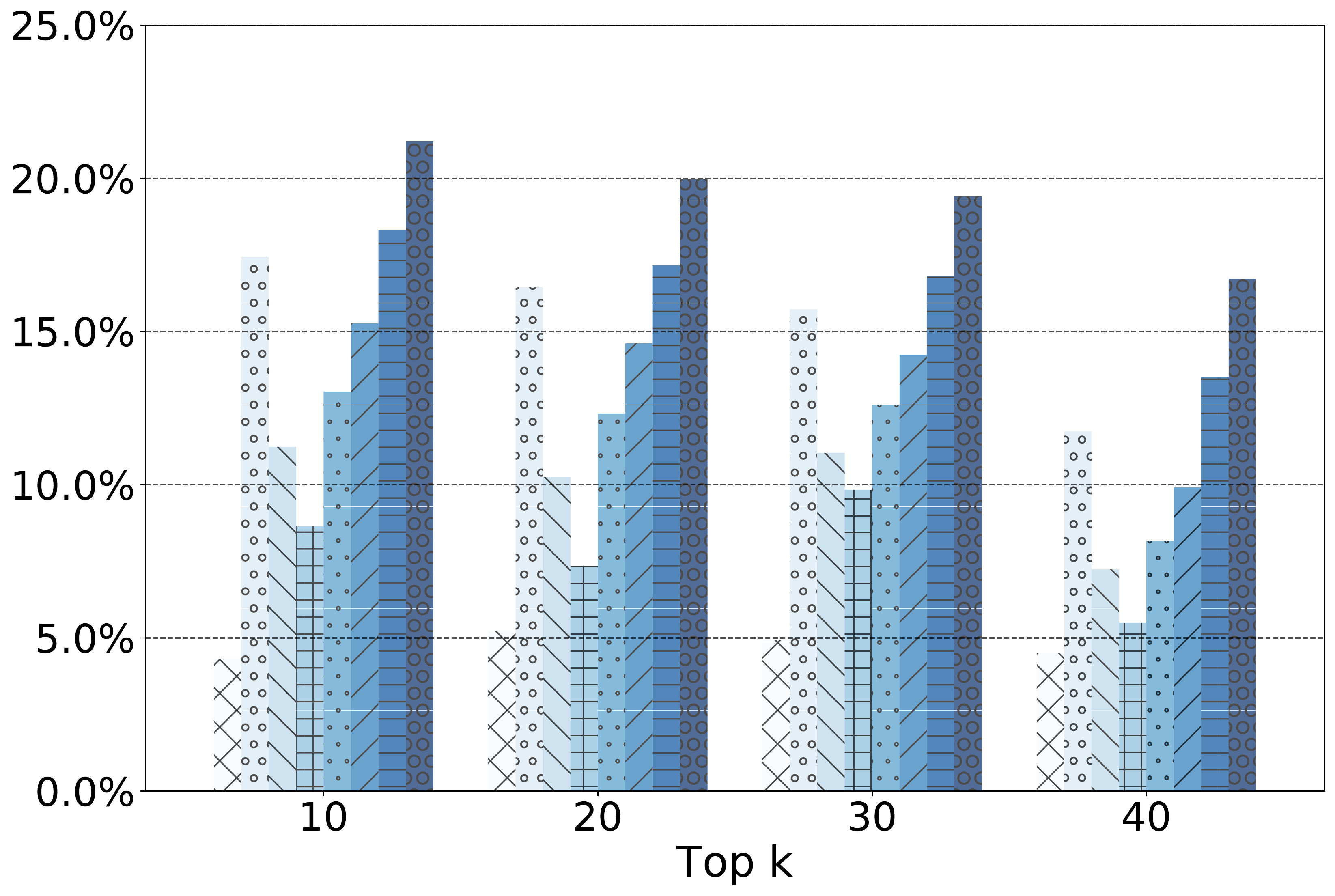}}
	\hspace{.05in} 
	\subfigure[The Sharpe Ratio on CSI 300.]{\includegraphics[width=0.49\columnwidth]{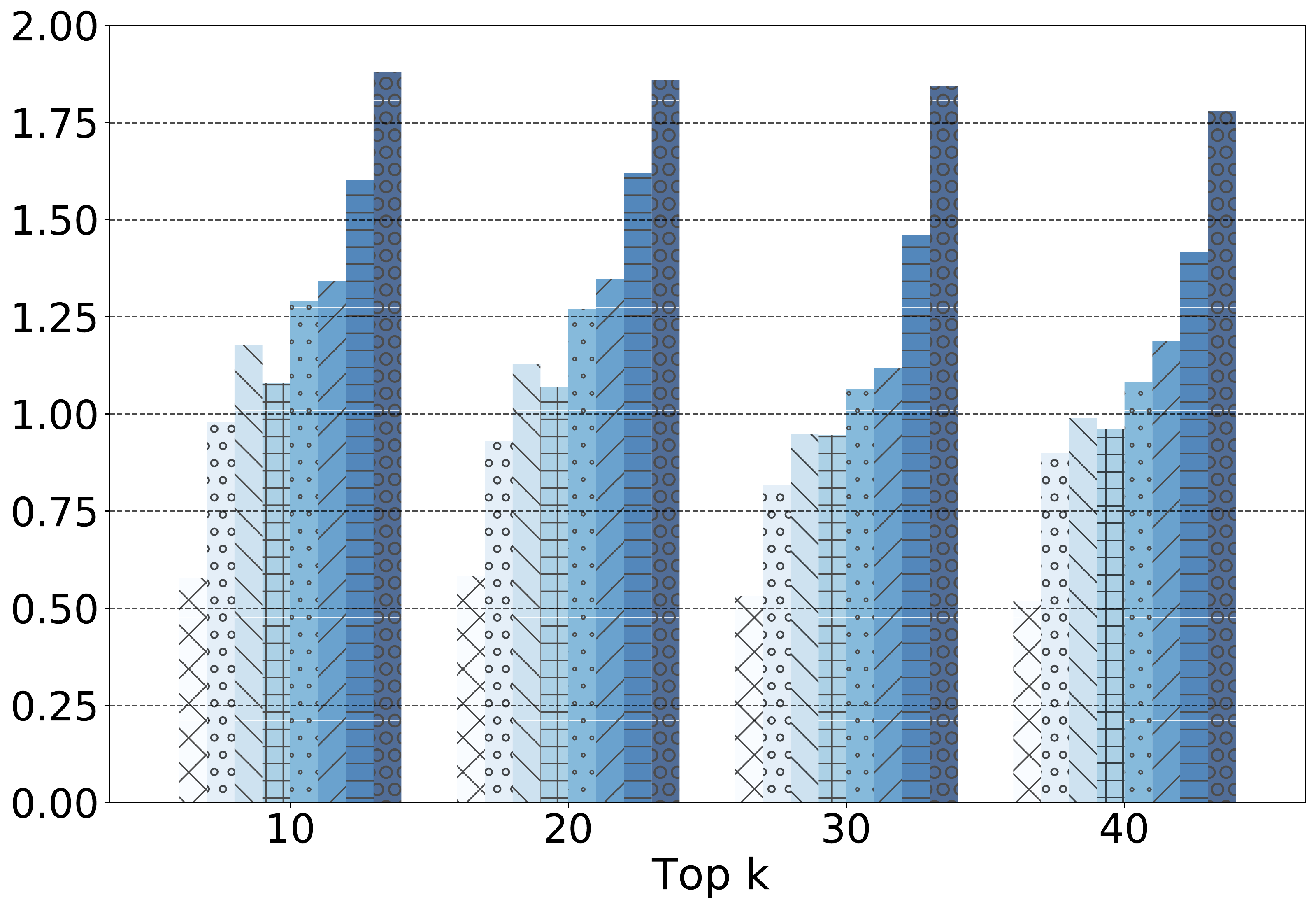}}
	\hspace{.05in} 
	\subfigure[The Sharpe Ratio on CSI 500.]{\includegraphics[width=0.49\columnwidth]{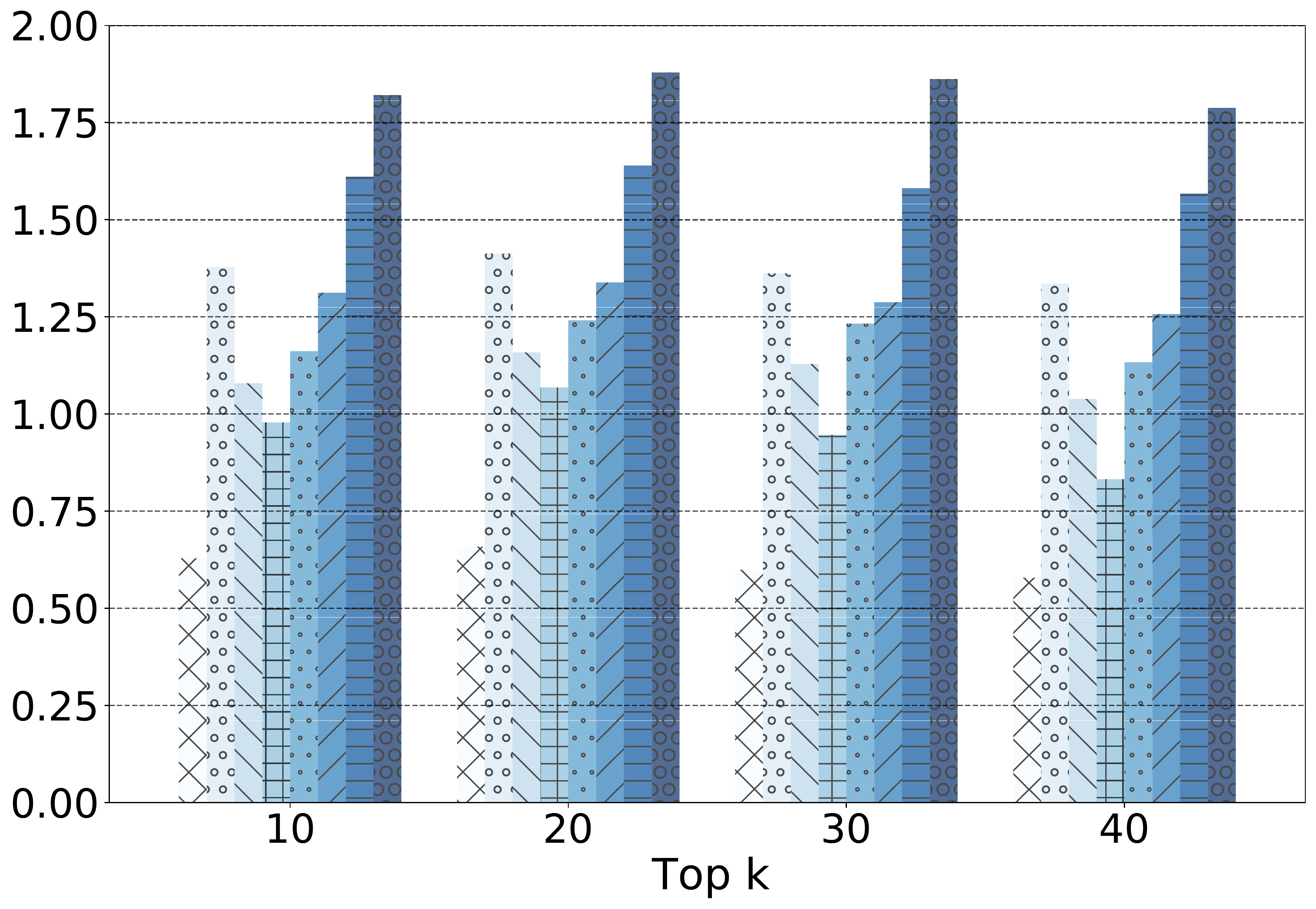}}
	\vspace{-.1in}
	\caption{The results of investment simulation.}
	\label{fig:simulation}
\end{figure*} 

\subsection{Experimental Setting}
\label{subsec:setting}
\subsubsection{Evaluation Metrics}
We use three standard evaluation metrics of regression to evaluate the results of stock trend forecasting, which are \textbf{Root Mean Squared Error (RMSE)}, \textbf{Mean Absolute Error (MAE)} and \textbf{Median Absolute Error (MedAE)}.
RMSE is calculated by: $\mathrm{RMSE}(y, \hat{y}) = \sqrt{\frac{1}{n} \sum_{i=1}^n (y_i - \hat{y}_i)^2}$;
MAE is the mean value of absolute error between labels and predictions: $\mathrm{MAE}(y, \hat{y}) = \frac{1}{n} \sum_{i=1}^n |y_i - \hat{y}_i|$;
MedAE is the median of absolute error between labels and predictions: $\mathrm{MedAE}(y, \hat{y}) = \mathrm{median}(|y_1-\hat{y}_1|, ..., |y_n - \hat{y}_n|)$. 
We repeat the testing procedure ten times for all the experimental results and report the average value to eliminate the fluctuations caused by different initialization.

\subsubsection{Hyper-parameters Setting}
In the type-specific text encoder, we set the dimension of type embeddings and token embeddings to $128$, and the number of head $K$ in multi-head attention is $4$; thus, the dimension of event embedding $e_i$ is $512$. The number of hidden units in the three LSTMs of our framework is $512$. We set the distance $l$ of event propagation on CSI 300 and CSI 500 to $3$ and $2$, respectively. The regularization parameter $\lambda$ is $2\times10^{-4}$ and the numbers of training epoch is $30$.

\subsubsection{Compared Methods} 
We compare our REST framework with the following stock trend forecasting methods:
\begin{itemize}[leftmargin=1.5em]
	\item \textbf{ARIMA}~\cite{brown2004smoothing}: ARIMA uses the historical stock trend's series of each stock as input to forecast the stock future trend.
	\item \textbf{TGCN}~\cite{matsunaga2019exploring}: TGCN leverages a variant of graph convolutional networks~\cite{kipf2017semi} to propagate the historical price information on the stock graph for stock trend forecasting. It only uses historical price data and does not use the event information.
	\item \textbf{HAN}~\cite{hu2018listening}: A model that utilizes the hierarchical attention mechanism over media information for stock trend forecasting.
	\item \textbf{Event-driven}: The Event-driven method directly feeds the original event information $\bar{H}^t$ in Section~\ref{subsec:learn_event} into a dense layer and output the prediction of the stock price trend. 
	\item \textbf{Event-driven\_sd}: The Event-driven\_sd method directly feeds the effect of event information $H_0^t$ in Section~\ref{subsec: learn_effect} into a dense layer and output the prediction of the stock price trend. The Event-driven\_sd method can learn the stock-dependent influence compared with the Event-driven method.
	
	\item \textbf{GCN}~\cite{kipf2017semi}: GCN model uses the GCN propagation layer in Section~\ref{subsubsec: GCN} to propagate the effect of event information on the stock graph.
	\item \textbf{Relational GCN}~\cite{lstm_rgcn}: This method uses the Relational GCN propagation layer in Section~\ref{subsubsec: RGCN} to propagate the effect of event information. It can learn the different event effects under different relations compared with the GCN model.
	\item \textbf{REST$_{l=1}$}: This method uses the dynamic propagation layer in Section~\ref{subsubsec: Dynamic_Propagation} to propagate the effect of event information on the stock graph. Nevertheless, it can not propagate the effect of event information through the multi-hop path. 
\end{itemize}

\begin{table}[t]
	\centering
	\caption{Experimental results on CSI 300 and CSI 500.}
	\resizebox{0.48\textwidth}{!}{
		\begin{tabular}{ l | c   c  c | c  c c}
			\hline
			\multirow{2}{*}{\textbf{Method}}  &  \multicolumn{3}{c|}{\textbf{CSI 300}} &   \multicolumn{3}{c}{\textbf{CSI 500}} \\
			\cmidrule{2-7}
			& \textbf{RMSE} & \textbf{MAE} & \textbf{MedAE}& \textbf{RMSE} & \textbf{MAE} & \textbf{MedAE}  \\ 
			\hline
			ARIMA~\cite{brown2004smoothing} &\cellcolor{blue!0}$0.0402$ &\cellcolor{blue!0}$0.0314$ &\cellcolor{blue!0}$0.0265$ &\cellcolor{blue!0}$0.0398$&\cellcolor{blue!0}$0.0311$ &\cellcolor{blue!0}$0.0264$ \\
			TGCN~\cite{matsunaga2019exploring} &\cellcolor{blue!10}$0.0387$ &\cellcolor{blue!10}$0.0301$ &\cellcolor{blue!10}$0.0250$ &\cellcolor{blue!35}$0.0347$  &\cellcolor{blue!35}$0.0276$ &\cellcolor{blue!35}$0.0215$\\
			HAN~\cite{hu2018listening}  &\cellcolor{blue!25}$0.0364$ &\cellcolor{blue!25}$0.0293$ &\cellcolor{blue!25}$0.0244$ &\cellcolor{blue!15}$0.0372$ &\cellcolor{blue!15}$0.0297$ &\cellcolor{blue!15}$0.0242$\\
			Event-driven &\cellcolor{blue!20}$0.0373$ &\cellcolor{blue!20}$0.0295$ &\cellcolor{blue!20}$0.0246$ &\cellcolor{blue!20}$0.0381$  &\cellcolor{blue!20}$0.0305$  &\cellcolor{blue!20}$0.0250$\\
			Event-driven\_sd &\cellcolor{blue!30}$0.0357$&\cellcolor{blue!30}$0.0284$ &\cellcolor{blue!30}$0.0238$&\cellcolor{blue!25}$0.0367$ &\cellcolor{blue!25}$0.0291$ &\cellcolor{blue!25}$0.0239$\\
			
			GCN~\cite{kipf2017semi}&\cellcolor{blue!35}$0.0344$ &\cellcolor{blue!35}$0.0267$ &\cellcolor{blue!35}$0.0225$ &\cellcolor{blue!30}$0.0358$ &\cellcolor{blue!30}$0.0284$ &\cellcolor{blue!30}$0.0226$\\
			Relational GCN~\cite{lstm_rgcn}&\cellcolor{blue!40}$0.0328$&\cellcolor{blue!40}$0.0246$ &\cellcolor{blue!40}$0.0207$ &\cellcolor{blue!40}$0.0339$ &\cellcolor{blue!40}$0.0266$ &\cellcolor{blue!40}$0.0204$\\
			REST$_{l=1}$&\cellcolor{blue!45}$0.0315$& \cellcolor{blue!45}$0.0240$ & \cellcolor{blue!45}$0.0196$&\cellcolor{blue!45}$0.0327$&\cellcolor{blue!45}$0.0254$& \cellcolor{blue!45}$0.0193$ \\
			\textbf{REST} &\cellcolor{blue!55}$\boldsymbol{0.0301}$ &\cellcolor{blue!55}$\boldsymbol{0.0227}$ &\cellcolor{blue!55}$\boldsymbol{0.0181}$&\cellcolor{blue!55}$\boldsymbol{0.0317}$ &\cellcolor{blue!55}$\boldsymbol{0.0242}$ &\cellcolor{blue!55}$\boldsymbol{0.0178}$\\
			\hline
	\end{tabular}}
	\label{tab:result_mse}
\end{table}

\subsection{Performance Comparison (RQ1)}
Table~\ref{tab:result_mse} shows the results of our REST framework and the compared methods on CSI 300 and CSI 500. 
Our REST framework can achieve the lowest RMSE, MAE, and MedAE compared with the previous methods. 
The Event-driven\_sd method is better than the Event-driven method, thus considering the stock-dependent influence of event information is essential.
The GCN model has better results than the Event-driven\_sd, which demonstrates that propagating the effect of event information can improve the performance of the stock trend forecasting. 
When we compare the GCN model and the Relational GCN model, we can find that learning the propagated event effect under different relations is vital. 
The comparison between the results of the Relational GCN model and REST$_{l=1}$ model shows that the dynamic propagated weights learned from the stock context are better than the fixed propagated weights. 
Finally, our REST's results are better than the REST$_{l=1}$ model, which verifies the effectiveness of the multi-hop propagation.

\subsection{Investment Simulation (RQ2)}
To further evaluate our framework's effectiveness, we conduct a trading strategy to simulate the stock investment. In detail, we rank the stocks on date $t$ from high to low according to their stock trend predictions, then select the top $k$ stocks to evenly invest and sell the holding shares of the stocks which rank behind $k$. We consider a transaction cost of $0.15\%$ for the buying shares and $0.25\%$ for the selling shares for approximating real-world trading. We use two financial metrics to evaluate the investment simulation result: the Annual Return and Sharpe Ratio~\cite{sharpe1966mutual}:

\begin{itemize}[leftmargin=1.5em]
	\item \textbf{Annual Return (AR)} is a standard profit indicator in finance, which is the return that an investment provides over a year.
	
	\item \textbf{Sharpe Ratio (SHR)} measures the performance of an investment compared to a risk-free asset, after adjusting for its risk. It is defined as: ${\rm SHR} = \dfrac{\rm E[R_a - R_b]}{\sqrt{\rm var[R_a - R_b]}}$, where $R_a$ is the return of given portfolio and $R_b$ denotes the risk-free return. $\rm E[\cdot]$ is the expected value, and $\rm var[\cdot]$ is the variance.
\end{itemize}
The number of stocks we select to invest evenly is $10$, $20$, $30$, and $40$, respectively. Figure~\ref{fig:simulation} shows the simulation results of the Annual Return and the Sharpe Ratio. Our REST framework can gain the highest Annual Return and Sharpe Ratio. The Event-driven\_sd method has a higher investment return than the Event-driven method, so the stock-dependent influence of event information is vital for stock trend forecasting. The GCN model has a higher investment return than the Event-driven\_sd model, which means the propagation of event information can make the forecasting more accurate. The investment results of Relational GCN show the effectiveness of learning the propagated event effect under different relations; the investment results of the REST$_{l=1}$ model demonstrate that dynamic propagated weights learned from the stock context can achieve a higher investment return; the comparison between the results of the REST$_{l=1}$ model and our REST framework verify the effectiveness of multi-hop propagation. 

\begin{table}[t]
	\centering
	\caption{Effect of the stock context.}
	\resizebox{0.48\textwidth}{!}{
		\begin{tabular}{ l  | c  c  c | c c c}
			\hline
			\textbf{}\multirow{2}{*}{\textbf{Stock Context}}  &  \multicolumn{3}{c|}{\textbf{CSI 300}} &   \multicolumn{3}{c}{\textbf{CSI 500}} \\
			\cmidrule{2-7}
			& \textbf{RMSE} & \textbf{MAE} & \textbf{MedAE} & \textbf{RMSE} & \textbf{MAE} & \textbf{MedAE}  \\ 
			\hline
			\rule{0pt}{10pt}
			REST$_{l=1}^{\mathrm{Event}}$ & \cellcolor{blue!10}$0.0326$& \cellcolor{blue!10}$0.0254$ & \cellcolor{blue!10}$0.0211$ & \cellcolor{blue!10}$0.0341$& \cellcolor{blue!10}$0.0265$ & \cellcolor{blue!10}$0.0208$\\
			\rule{0pt}{10pt}
			REST$_{l=1}^{\mathrm{Feedback}}$ &\cellcolor{blue!20}$0.0321$&\cellcolor{blue!20}$0.0245$ & \cellcolor{blue!20}$0.0204$  &\cellcolor{blue!20}$0.0334$&\cellcolor{blue!20}$0.0259$ & \cellcolor{blue!20}$0.0204$\\
			\rule{0pt}{10pt}
			REST$_{l=1}$& \cellcolor{blue!30}$\boldsymbol{0.0315}$& \cellcolor{blue!30}$\boldsymbol{0.0240}$ &\cellcolor{blue!30}$\boldsymbol{0.0196}$ & \cellcolor{blue!30}$\boldsymbol{0.0327}$& \cellcolor{blue!30}$\boldsymbol{0.0254}$ &\cellcolor{blue!30}$\boldsymbol{0.0193}$ \\
			\hline
	\end{tabular}}
	\label{tab:effect_state}
\end{table}
\begin{table}[t]
	\centering
	\caption{ Effect of the distance of event propagation.}
	\resizebox{0.48\textwidth}{!}{
		\begin{tabular}{ l  | c  c c | c c c}
			\hline
			\multirow{2}{*}{\textbf{Distance}} &  \multicolumn{3}{c|}{\textbf{CSI 300}} &   \multicolumn{3}{c}{\textbf{CSI 500}} \\
			\cmidrule{2-7}
			& \textbf{RMSE} & \textbf{MAE} & \textbf{MedAE} & \textbf{RMSE} & \textbf{MAE} & \textbf{MedAE}  \\ 
			\hline
			REST$_{l=1}$ & \cellcolor{blue!20}$0.0315$&\cellcolor{blue!20}$0.0240$ &\cellcolor{blue!20}$0.0196$ & \cellcolor{blue!20}$0.0327$&\cellcolor{blue!20}$0.0254$ &\cellcolor{blue!20}$0.0193$ \\
			
			REST$_{l=2}$ &\cellcolor{blue!45}$0.0308$ & \cellcolor{blue!45}$0.0234$ &\cellcolor{blue!45}$0.0185$ &\cellcolor{blue!45}$0.0317$ & \cellcolor{blue!55}$\boldsymbol{0.0242}$ &\cellcolor{blue!55}$\boldsymbol{0.0178}$ \\
			
			REST$_{l=3}$ & \cellcolor{blue!55}$\boldsymbol{0.0301}$ &  \cellcolor{blue!55}$\boldsymbol{0.0227}$ & \cellcolor{blue!55}$\boldsymbol{0.0181}$ & \cellcolor{blue!55}$\boldsymbol{0.0314}$ &\cellcolor{blue!45}$0.0244$ & \cellcolor{blue!45}$0.0181$\\
			
			REST$_{l=4}$ & \cellcolor{blue!35}$0.0312$ & \cellcolor{blue!35}$0.0237$ & \cellcolor{blue!35}$0.0190$ & \cellcolor{blue!35}$0.0320$ & \cellcolor{blue!35}$0.0251$ & \cellcolor{blue!35}$0.0186$  \\
			\hline
	\end{tabular}}
	\label{tab:effect_dist}
\end{table} 
\subsection{Model Analysis}
\subsubsection{Effect of the Stock Context (RQ3)}
\label{sec:effect_state}
To explore the effect of the stock context described in Section~\ref{subsec: learn_state}. We observe the performance of REST$_{l=1}$'s variants: 1) REST$_{l=1}^{\mathrm{Event}}$ merely uses the historical events to learn the stock context; 2) REST$_{l=1}^{\mathrm{Feedback}}$ only uses the historical events' feedbacks to learn the stock context.
We summarize the results in Table~\ref{tab:effect_state}, and have the following observations:
\begin{itemize}[leftmargin=1.5em]
	\item  From the comparison between REST$_{l=1}^{\mathrm{Event}}$ and REST$_{l=1}^{\mathrm{Feedback}}$, we can find that using historical events' feedbacks to learn the stock context is better than historical events. It verifies the effectiveness of the historical events' feedbacks in the stock trend forecasting.
	\item Using historical events and historical events' feedbacks simultaneously can achieve the best performance. Thus we need to jointly use historical events and historical events' feedbacks to learn the stock context.
\end{itemize}

\subsubsection{Effect of the Distance of Event Propagation (RQ3)}
\label{sec:effect_dist}
To study the effect of multi-hop propagation's distance $l$ in Section~\ref{subsubsec:multi-hop}, we vary the number of propagation layer of the REST framework from $1$ to $4$ and observe the results on CSI 300 and CSI 500. 
Table~\ref{tab:effect_dist} summarizes the experiments. 
Then we have the following observations:

\begin{itemize}[leftmargin=1.5em]
	\item  Increasing the distance $l$ of propagation is capable of boosting the performance substantially. When $l=2$, the REST$_{l=2}$ achieves the lowest MAE and MedAE on CSI 500; when $l=3$, the REST$_{l=3}$ achieves the best RMSE, MAE and MedAE on CSI 300, and lowest RMSE on CSI 500.
	\item When we further increase the number of propagation layers to $4$, we observe that our framework's performance will reduce. It suggests that $l$ is a sensitive hyper-parameter, and too much of the propagation layer can not further improve stock trend forecasting.
\end{itemize}

\begin{figure}[t]
	\centering
	\includegraphics[width=0.98\columnwidth]{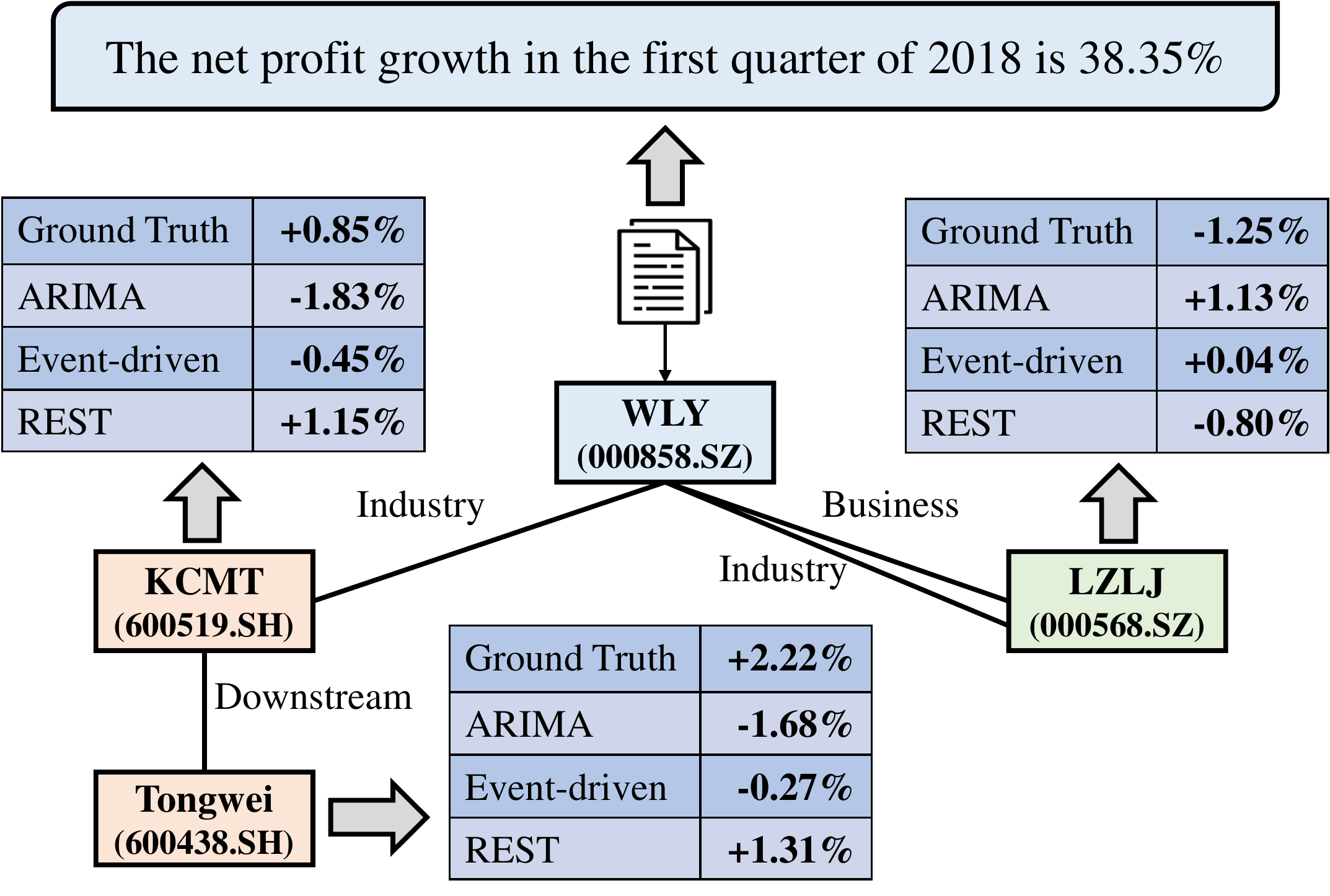}
	\caption{The ground truth of stock price trend and the forecasting results of ARIMA, Event-driven and REST methods.}
	\label{fig:casestudy}
\end{figure}
\subsubsection{Case Study (RQ4)}
To illustrate our REST framework's effectiveness, we dig into a specific case study on three companies: \textbf{WLY}, \textbf{KCMT}, and \textbf{LZLJ}. While belonging to the same general industry, these three stocks connect yield complicated competitive relations, making it quite hard to discern the influence of one company's events on the others.

For example, on April 28, 2018, the stock \textbf{WLY} released a performance announcement that the net profit growth in the first quarter of 2018 is 38.35\%.
Such a positive event of \textbf{WLY} can intuitively imply a bullish trend of the corresponding industry. However, complicated competition between different companies can lead to diverse stories. Particularly, while the stock \textbf{KCMT}, whose primary products are distinct as \textbf{WLY}, can benefit from this positive event, another stock, \textbf{LZLJ}, having the same type of product thus the more competitive position as \textbf{WLY}, could suffer from the adverse effects of the positive event of \textbf{WLY}. 

We observe the forecasting results of different methods on the closest trading day after this event happened.
Figure~\ref{fig:casestudy} illustrates the ground truth of the stock price trend and the forecasting results of ARIMA, Event-driven, and REST. 
The ARIMA forecasts the stock trend with historical stock trend series only, and the Event-driven method can not learn the cross-stock influence of event information, so they both fail to make the correct forecastings. 
However, our REST framework can propagate the effect of event information with the propagation layer we designed. Thus, our REST framework can make the correct forecasting on the \textbf{KCMT} and \textbf{LZLJ}. 
The effect of \textbf{WLY}'s event on other related stock is difficult to be found by ordinary people unless experts with rich domain knowledge. Our REST framework can make the correct numerical prediction while the experts can only forecast a rise or a fall.

Moreover, our REST framework can propagate the event information through multi-hop paths, so our framework can also make correct forecasting on the stock \textbf{Tongwei}, which is the downstream company of \textbf{KCMT}.

%% file: conclusion.tex
\section{Conclusion and Future Work}
\label{sec:conclusion}
Stock trend forecasting is vital for investors to seek maximized profits from financial investment. 
In this paper, we proposed a relational event-driven framework (REST) to forecast the stock trend. Our proposed framework can address some shortcomings of existing event-driven work. The first shortcoming is existing methods overlook the influence of event effect differentiated by the stock-dependent properties (stock-dependent influence). The second one is that existing methods only use one stock's information to predict its stock price trend, overlooking the effect of event information from other different stocks on the stock trend of this stock (cross-stock influence). 
The experiment results on real-world stock market data demonstrated the effectiveness of our framework. The results of the investment simulation show that our REST framework can achieve a higher investment return.

In the future, we would mine more abundant and diverse stock information from the Web, such as the news, social media, and other stock related information, and fuse different kinds of stock information to forecasting the stock price trend. We would also explore the techniques of unsupervised learning and apply unsupervised learning to stock trend forecasting.

\section*{Acknowledgments}
Wentao Xu and Jian Yin are supported by the National Natural Science Foundation of China (U1711262, U1711261,U1811264,U1811261,\\U1911203,U2001211), Guangdong Basic and Applied Basic Research Foundation (2019B1515130001),  Key R\&D Program of Guangdong Province (2018B010107005). Jian Yin is the corresponding author.

%% file: main.bbl

\begin{thebibliography}{46}


\ifx \showCODEN    \undefined \def \showCODEN     #1{\unskip}     \fi
\ifx \showDOI      \undefined \def \showDOI       #1{#1}\fi
\ifx \showISBNx    \undefined \def \showISBNx     #1{\unskip}     \fi
\ifx \showISBNxiii \undefined \def \showISBNxiii  #1{\unskip}     \fi
\ifx \showISSN     \undefined \def \showISSN      #1{\unskip}     \fi
\ifx \showLCCN     \undefined \def \showLCCN      #1{\unskip}     \fi
\ifx \shownote     \undefined \def \shownote      #1{#1}          \fi
\ifx \showarticletitle \undefined \def \showarticletitle #1{#1}   \fi
\ifx \showURL      \undefined \def \showURL       {\relax}        \fi
\providecommand\bibfield[2]{#2}
\providecommand\bibinfo[2]{#2}
\providecommand\natexlab[1]{#1}
\providecommand\showeprint[2][]{arXiv:#2}

\bibitem[\protect\citeauthoryear{Akita, Yoshihara, Matsubara, and Uehara}{Akita
  et~al\mbox{.}}{2016}]%
        {akita2016deep}
\bibfield{author}{\bibinfo{person}{Ryo Akita}, \bibinfo{person}{Akira
  Yoshihara}, \bibinfo{person}{Takashi Matsubara}, {and}
  \bibinfo{person}{Kuniaki Uehara}.} \bibinfo{year}{2016}\natexlab{}.
\newblock \showarticletitle{Deep learning for stock prediction using numerical
  and textual information}. In \bibinfo{booktitle}{\emph{2016 IEEE/ACIS 15th
  International Conference on Computer and Information Science (ICIS)}}. IEEE,
  \bibinfo{pages}{1--6}.
\newblock


\bibitem[\protect\citeauthoryear{Ariyo, Adewumi, and Ayo}{Ariyo
  et~al\mbox{.}}{2014}]%
        {ariyo2014stock}
\bibfield{author}{\bibinfo{person}{Adebiyi~A Ariyo}, \bibinfo{person}{Adewumi~O
  Adewumi}, {and} \bibinfo{person}{Charles~K Ayo}.}
  \bibinfo{year}{2014}\natexlab{}.
\newblock \showarticletitle{Stock price prediction using the ARIMA model}. In
  \bibinfo{booktitle}{\emph{2014 UKSim-AMSS 16th International Conference on
  Computer Modelling and Simulation}}. IEEE, \bibinfo{pages}{106--112}.
\newblock


\bibitem[\protect\citeauthoryear{Bao, Yue, and Rao}{Bao et~al\mbox{.}}{2017}]%
        {bao2017deep}
\bibfield{author}{\bibinfo{person}{Wei Bao}, \bibinfo{person}{Jun Yue}, {and}
  \bibinfo{person}{Yulei Rao}.} \bibinfo{year}{2017}\natexlab{}.
\newblock \showarticletitle{A deep learning framework for financial time series
  using stacked autoencoders and long-short term memory}.
\newblock \bibinfo{journal}{\emph{PloS one}} \bibinfo{volume}{12},
  \bibinfo{number}{7} (\bibinfo{year}{2017}), \bibinfo{pages}{e0180944}.
\newblock


\bibitem[\protect\citeauthoryear{Berkowitz, Logue, and Noser~Jr}{Berkowitz
  et~al\mbox{.}}{1988}]%
        {berkowitz1988total}
\bibfield{author}{\bibinfo{person}{Stephen~A Berkowitz},
  \bibinfo{person}{Dennis~E Logue}, {and} \bibinfo{person}{Eugene~A Noser~Jr}.}
  \bibinfo{year}{1988}\natexlab{}.
\newblock \showarticletitle{The total cost of transactions on the NYSE}.
\newblock \bibinfo{journal}{\emph{The Journal of Finance}}
  \bibinfo{volume}{43}, \bibinfo{number}{1} (\bibinfo{year}{1988}),
  \bibinfo{pages}{97--112}.
\newblock


\bibitem[\protect\citeauthoryear{Brown}{Brown}{2004}]%
        {brown2004smoothing}
\bibfield{author}{\bibinfo{person}{Robert~Goodell Brown}.}
  \bibinfo{year}{2004}\natexlab{}.
\newblock \bibinfo{booktitle}{\emph{Smoothing, forecasting and prediction of
  discrete time series}}.
\newblock \bibinfo{publisher}{Courier Corporation}.
\newblock


\bibitem[\protect\citeauthoryear{Chen, Wei, and Huang}{Chen
  et~al\mbox{.}}{2018}]%
        {chen2018incorporating}
\bibfield{author}{\bibinfo{person}{Yingmei Chen}, \bibinfo{person}{Zhongyu
  Wei}, {and} \bibinfo{person}{Xuanjing Huang}.}
  \bibinfo{year}{2018}\natexlab{}.
\newblock \showarticletitle{Incorporating Corporation Relationship via Graph
  Convolutional Neural Networks for Stock Price Prediction}. In
  \bibinfo{booktitle}{\emph{Proceedings of the 27th ACM International
  Conference on Information and Knowledge Management}}. ACM,
  \bibinfo{pages}{1655--1658}.
\newblock


\bibitem[\protect\citeauthoryear{Deng, Zhang, Zhang, Chen, Pan, and Chen}{Deng
  et~al\mbox{.}}{2019}]%
        {deng2019knowledge}
\bibfield{author}{\bibinfo{person}{Shumin Deng}, \bibinfo{person}{Ningyu
  Zhang}, \bibinfo{person}{Wen Zhang}, \bibinfo{person}{Jiaoyan Chen},
  \bibinfo{person}{Jeff~Z Pan}, {and} \bibinfo{person}{Huajun Chen}.}
  \bibinfo{year}{2019}\natexlab{}.
\newblock \showarticletitle{Knowledge-Driven Stock Trend Prediction and
  Explanation via Temporal Convolutional Network}. In
  \bibinfo{booktitle}{\emph{Companion Proceedings of The 2019 World Wide Web
  Conference}}. ACM, \bibinfo{pages}{678--685}.
\newblock


\bibitem[\protect\citeauthoryear{Ding, Zhang, Liu, and Duan}{Ding
  et~al\mbox{.}}{2016}]%
        {ding2016knowledge}
\bibfield{author}{\bibinfo{person}{Xiao Ding}, \bibinfo{person}{Yue Zhang},
  \bibinfo{person}{Ting Liu}, {and} \bibinfo{person}{Junwen Duan}.}
  \bibinfo{year}{2016}\natexlab{}.
\newblock \showarticletitle{Knowledge-driven event embedding for stock
  prediction}. In \bibinfo{booktitle}{\emph{Proceedings of coling 2016, the
  26th international conference on computational linguistics: Technical
  papers}}. \bibinfo{pages}{2133--2142}.
\newblock


\bibitem[\protect\citeauthoryear{Duan, Zhang, Ding, Chang, and Liu}{Duan
  et~al\mbox{.}}{2018}]%
        {duan2018learning}
\bibfield{author}{\bibinfo{person}{Junwen Duan}, \bibinfo{person}{Yue Zhang},
  \bibinfo{person}{Xiao Ding}, \bibinfo{person}{Ching-Yun Chang}, {and}
  \bibinfo{person}{Ting Liu}.} \bibinfo{year}{2018}\natexlab{}.
\newblock \showarticletitle{Learning Target-Specific Representations of
  Financial News Documents For Cumulative Abnormal Return Prediction}. In
  \bibinfo{booktitle}{\emph{Proceedings of the 27th International Conference on
  Computational Linguistics}}. \bibinfo{pages}{2823--2833}.
\newblock


\bibitem[\protect\citeauthoryear{Edwards, Magee, and Bassetti}{Edwards
  et~al\mbox{.}}{2018}]%
        {edwards2018technical}
\bibfield{author}{\bibinfo{person}{Robert~D Edwards}, \bibinfo{person}{John
  Magee}, {and} \bibinfo{person}{WH~Charles Bassetti}.}
  \bibinfo{year}{2018}\natexlab{}.
\newblock \bibinfo{booktitle}{\emph{Technical analysis of stock trends}}.
\newblock \bibinfo{publisher}{CRC press}.
\newblock


\bibitem[\protect\citeauthoryear{Feng, Chen, He, Ding, Sun, and Chua}{Feng
  et~al\mbox{.}}{2019a}]%
        {feng2019enhancing}
\bibfield{author}{\bibinfo{person}{Fuli Feng}, \bibinfo{person}{Huimin Chen},
  \bibinfo{person}{Xiangnan He}, \bibinfo{person}{Ji Ding},
  \bibinfo{person}{Maosong Sun}, {and} \bibinfo{person}{Tat-Seng Chua}.}
  \bibinfo{year}{2019}\natexlab{a}.
\newblock \showarticletitle{Enhancing Stock Movement Prediction with
  Adversarial Training}.
\newblock \bibinfo{journal}{\emph{IJCAI}} (\bibinfo{year}{2019}).
\newblock


\bibitem[\protect\citeauthoryear{Feng, He, Wang, Luo, Liu, and Chua}{Feng
  et~al\mbox{.}}{2019b}]%
        {feng2019temporal}
\bibfield{author}{\bibinfo{person}{Fuli Feng}, \bibinfo{person}{Xiangnan He},
  \bibinfo{person}{Xiang Wang}, \bibinfo{person}{Cheng Luo},
  \bibinfo{person}{Yiqun Liu}, {and} \bibinfo{person}{Tat-Seng Chua}.}
  \bibinfo{year}{2019}\natexlab{b}.
\newblock \showarticletitle{Temporal relational ranking for stock prediction}.
\newblock \bibinfo{journal}{\emph{ACM Transactions on Information Systems
  (TOIS)}} \bibinfo{volume}{37}, \bibinfo{number}{2} (\bibinfo{year}{2019}),
  \bibinfo{pages}{27}.
\newblock


\bibitem[\protect\citeauthoryear{Gao}{Gao}{2016}]%
        {gao2016stock}
\bibfield{author}{\bibinfo{person}{Qiyuan Gao}.}
  \bibinfo{year}{2016}\natexlab{}.
\newblock \emph{\bibinfo{title}{Stock market forecasting using recurrent neural
  network}}.
\newblock \bibinfo{thesistype}{Ph.D. Dissertation}. \bibinfo{school}{University
  of Missouri--Columbia}.
\newblock


\bibitem[\protect\citeauthoryear{Hochreiter and Schmidhuber}{Hochreiter and
  Schmidhuber}{1997}]%
        {hochreiter1997long}
\bibfield{author}{\bibinfo{person}{Sepp Hochreiter} {and}
  \bibinfo{person}{J{\"u}rgen Schmidhuber}.} \bibinfo{year}{1997}\natexlab{}.
\newblock \showarticletitle{Long short-term memory}.
\newblock \bibinfo{journal}{\emph{Neural computation}} \bibinfo{volume}{9},
  \bibinfo{number}{8} (\bibinfo{year}{1997}), \bibinfo{pages}{1735--1780}.
\newblock


\bibitem[\protect\citeauthoryear{Hu, Liu, Bian, Liu, and Liu}{Hu
  et~al\mbox{.}}{2018}]%
        {hu2018listening}
\bibfield{author}{\bibinfo{person}{Ziniu Hu}, \bibinfo{person}{Weiqing Liu},
  \bibinfo{person}{Jiang Bian}, \bibinfo{person}{Xuanzhe Liu}, {and}
  \bibinfo{person}{Tie-Yan Liu}.} \bibinfo{year}{2018}\natexlab{}.
\newblock \showarticletitle{Listening to chaotic whispers: A deep learning
  framework for news-oriented stock trend prediction}. In
  \bibinfo{booktitle}{\emph{Proceedings of the Eleventh ACM International
  Conference on Web Search and Data Mining}}. ACM, \bibinfo{pages}{261--269}.
\newblock


\bibitem[\protect\citeauthoryear{Kim, So, Jeong, Lee, Kim, and Kang}{Kim
  et~al\mbox{.}}{2019}]%
        {kim2019hats}
\bibfield{author}{\bibinfo{person}{Raehyun Kim}, \bibinfo{person}{Chan~Ho So},
  \bibinfo{person}{Minbyul Jeong}, \bibinfo{person}{Sanghoon Lee},
  \bibinfo{person}{Jinkyu Kim}, {and} \bibinfo{person}{Jaewoo Kang}.}
  \bibinfo{year}{2019}\natexlab{}.
\newblock \showarticletitle{Hats: A hierarchical graph attention network for
  stock movement prediction}.
\newblock \bibinfo{journal}{\emph{arXiv preprint arXiv:1908.07999}}
  (\bibinfo{year}{2019}).
\newblock


\bibitem[\protect\citeauthoryear{Kipf and Welling}{Kipf and Welling}{2017}]%
        {kipf2017semi}
\bibfield{author}{\bibinfo{person}{Thomas~N. Kipf} {and} \bibinfo{person}{Max
  Welling}.} \bibinfo{year}{2017}\natexlab{}.
\newblock \showarticletitle{Semi-Supervised Classification with Graph
  Convolutional Networks}. In \bibinfo{booktitle}{\emph{International
  Conference on Learning Representations (ICLR)}}.
\newblock


\bibitem[\protect\citeauthoryear{Li, Song, and Tao}{Li et~al\mbox{.}}{2019}]%
        {li2019multi}
\bibfield{author}{\bibinfo{person}{Chang Li}, \bibinfo{person}{Dongjin Song},
  {and} \bibinfo{person}{Dacheng Tao}.} \bibinfo{year}{2019}\natexlab{}.
\newblock \showarticletitle{Multi-task Recurrent Neural Networks and
  Higher-order Markov Random Fields for Stock Price Movement Prediction:
  Multi-task RNN and Higer-order MRFs for Stock Price Classification}. In
  \bibinfo{booktitle}{\emph{Proceedings of the 25th ACM SIGKDD International
  Conference on Knowledge Discovery \& Data Mining}}. ACM,
  \bibinfo{pages}{1141--1151}.
\newblock


\bibitem[\protect\citeauthoryear{Li, Leng, Yang, and Yu}{Li
  et~al\mbox{.}}{2016}]%
        {li2016stock}
\bibfield{author}{\bibinfo{person}{Lili Li}, \bibinfo{person}{Shan Leng},
  \bibinfo{person}{Jun Yang}, {and} \bibinfo{person}{Mei Yu}.}
  \bibinfo{year}{2016}\natexlab{}.
\newblock \showarticletitle{Stock Market Autoregressive Dynamics: A
  Multinational Comparative Study with Quantile Regression}.
\newblock \bibinfo{journal}{\emph{Mathematical Problems in Engineering}}
  \bibinfo{volume}{2016} (\bibinfo{year}{2016}).
\newblock


\bibitem[\protect\citeauthoryear{Li, Chen, Wang, Chen, and Chen}{Li
  et~al\mbox{.}}{2017}]%
        {li2017web}
\bibfield{author}{\bibinfo{person}{Qing Li}, \bibinfo{person}{Yan Chen},
  \bibinfo{person}{Jun Wang}, \bibinfo{person}{Yuanzhu Chen}, {and}
  \bibinfo{person}{Hsinchun Chen}.} \bibinfo{year}{2017}\natexlab{}.
\newblock \showarticletitle{Web media and stock markets: A survey and future
  directions from a big data perspective}.
\newblock \bibinfo{journal}{\emph{IEEE Transactions on Knowledge and Data
  Engineering}} \bibinfo{volume}{30}, \bibinfo{number}{2}
  (\bibinfo{year}{2017}), \bibinfo{pages}{381--399}.
\newblock


\bibitem[\protect\citeauthoryear{Li, Wang, Li, Liu, Gong, and Chen}{Li
  et~al\mbox{.}}{2014}]%
        {li2014effect}
\bibfield{author}{\bibinfo{person}{Qing Li}, \bibinfo{person}{TieJun Wang},
  \bibinfo{person}{Ping Li}, \bibinfo{person}{Ling Liu}, \bibinfo{person}{Qixu
  Gong}, {and} \bibinfo{person}{Yuanzhu Chen}.}
  \bibinfo{year}{2014}\natexlab{}.
\newblock \showarticletitle{The effect of news and public mood on stock
  movements}.
\newblock \bibinfo{journal}{\emph{Information Sciences}}  \bibinfo{volume}{278}
  (\bibinfo{year}{2014}), \bibinfo{pages}{826--840}.
\newblock


\bibitem[\protect\citeauthoryear{Li, Bao, Harimoto, Chen, Xu, and Su}{Li
  et~al\mbox{.}}{2020}]%
        {lstm_rgcn}
\bibfield{author}{\bibinfo{person}{Wei Li}, \bibinfo{person}{Ruihan Bao},
  \bibinfo{person}{Keiko Harimoto}, \bibinfo{person}{Deli Chen},
  \bibinfo{person}{Jingjing Xu}, {and} \bibinfo{person}{Qi Su}.}
  \bibinfo{year}{2020}\natexlab{}.
\newblock \showarticletitle{Modeling the Stock Relation with Graph Network for
  Overnight Stock Movement Prediction}. \bibinfo{pages}{4491--4497}.
\newblock


\bibitem[\protect\citeauthoryear{Liu, Cheng, Su, and Zhu}{Liu
  et~al\mbox{.}}{2018}]%
        {DBLP:conf/cikm/Liu0SZ18}
\bibfield{author}{\bibinfo{person}{Qikai Liu}, \bibinfo{person}{Xiang Cheng},
  \bibinfo{person}{Sen Su}, {and} \bibinfo{person}{Shuguang Zhu}.}
  \bibinfo{year}{2018}\natexlab{}.
\newblock \showarticletitle{Hierarchical Complementary Attention Network for
  Predicting Stock Price Movements with News}. In
  \bibinfo{booktitle}{\emph{Proceedings of the 27th {ACM} International
  Conference on Information and Knowledge Management, {CIKM} 2018, Torino,
  Italy, October 22-26, 2018}}. \bibinfo{pages}{1603--1606}.
\newblock
\urldef\tempurl%
\url{https://doi.org/10.1145/3269206.3269286}
\showDOI{\tempurl}


\bibitem[\protect\citeauthoryear{Malkiel and Fama}{Malkiel and Fama}{1970}]%
        {malkiel1970efficient}
\bibfield{author}{\bibinfo{person}{Burton~G Malkiel} {and}
  \bibinfo{person}{Eugene~F Fama}.} \bibinfo{year}{1970}\natexlab{}.
\newblock \showarticletitle{Efficient capital markets: A review of theory and
  empirical work}.
\newblock \bibinfo{journal}{\emph{The journal of Finance}}
  \bibinfo{volume}{25}, \bibinfo{number}{2} (\bibinfo{year}{1970}),
  \bibinfo{pages}{383--417}.
\newblock


\bibitem[\protect\citeauthoryear{Matsunaga, Suzumura, and Takahashi}{Matsunaga
  et~al\mbox{.}}{2019}]%
        {matsunaga2019exploring}
\bibfield{author}{\bibinfo{person}{Daiki Matsunaga}, \bibinfo{person}{Toyotaro
  Suzumura}, {and} \bibinfo{person}{Toshihiro Takahashi}.}
  \bibinfo{year}{2019}\natexlab{}.
\newblock \showarticletitle{Exploring Graph Neural Networks for Stock Market
  Predictions with Rolling Window Analysis}.
\newblock \bibinfo{journal}{\emph{arXiv preprint arXiv:1909.10660}}
  (\bibinfo{year}{2019}).
\newblock


\bibitem[\protect\citeauthoryear{Nassirtoussi, Aghabozorgi, Wah, and
  Ngo}{Nassirtoussi et~al\mbox{.}}{2015}]%
        {nassirtoussi2015text}
\bibfield{author}{\bibinfo{person}{Arman~Khadjeh Nassirtoussi},
  \bibinfo{person}{Saeed Aghabozorgi}, \bibinfo{person}{Teh~Ying Wah}, {and}
  \bibinfo{person}{David Chek~Ling Ngo}.} \bibinfo{year}{2015}\natexlab{}.
\newblock \showarticletitle{Text mining of news-headlines for FOREX market
  prediction: A Multi-layer Dimension Reduction Algorithm with semantics and
  sentiment}.
\newblock \bibinfo{journal}{\emph{Expert Systems with Applications}}
  \bibinfo{volume}{42}, \bibinfo{number}{1} (\bibinfo{year}{2015}),
  \bibinfo{pages}{306--324}.
\newblock


\bibitem[\protect\citeauthoryear{Nguyen, Shirai, and Velcin}{Nguyen
  et~al\mbox{.}}{2015}]%
        {nguyen2015sentiment}
\bibfield{author}{\bibinfo{person}{Thien~Hai Nguyen}, \bibinfo{person}{Kiyoaki
  Shirai}, {and} \bibinfo{person}{Julien Velcin}.}
  \bibinfo{year}{2015}\natexlab{}.
\newblock \showarticletitle{Sentiment analysis on social media for stock
  movement prediction}.
\newblock \bibinfo{journal}{\emph{Expert Systems with Applications}}
  \bibinfo{volume}{42}, \bibinfo{number}{24} (\bibinfo{year}{2015}),
  \bibinfo{pages}{9603--9611}.
\newblock


\bibitem[\protect\citeauthoryear{Patel, Shah, Thakkar, and Kotecha}{Patel
  et~al\mbox{.}}{2015}]%
        {patel2015predicting}
\bibfield{author}{\bibinfo{person}{Jigar Patel}, \bibinfo{person}{Sahil Shah},
  \bibinfo{person}{Priyank Thakkar}, {and} \bibinfo{person}{Ketan Kotecha}.}
  \bibinfo{year}{2015}\natexlab{}.
\newblock \showarticletitle{Predicting stock market index using fusion of
  machine learning techniques}.
\newblock \bibinfo{journal}{\emph{Expert Systems with Applications}}
  \bibinfo{volume}{42}, \bibinfo{number}{4} (\bibinfo{year}{2015}),
  \bibinfo{pages}{2162--2172}.
\newblock


\bibitem[\protect\citeauthoryear{Qin and Yang}{Qin and Yang}{2019}]%
        {qin2019you}
\bibfield{author}{\bibinfo{person}{Yu Qin} {and} \bibinfo{person}{Yi Yang}.}
  \bibinfo{year}{2019}\natexlab{}.
\newblock \showarticletitle{What you say and how you say it matters: Predicting
  stock volatility using verbal and vocal cues}. In
  \bibinfo{booktitle}{\emph{Proceedings of the 57th Annual Meeting of the
  Association for Computational Linguistics}}. \bibinfo{pages}{390--401}.
\newblock


\bibitem[\protect\citeauthoryear{Rather, Agarwal, and Sastry}{Rather
  et~al\mbox{.}}{2015}]%
        {rather2015recurrent}
\bibfield{author}{\bibinfo{person}{Akhter~Mohiuddin Rather},
  \bibinfo{person}{Arun Agarwal}, {and} \bibinfo{person}{VN Sastry}.}
  \bibinfo{year}{2015}\natexlab{}.
\newblock \showarticletitle{Recurrent neural network and a hybrid model for
  prediction of stock returns}.
\newblock \bibinfo{journal}{\emph{Expert Systems with Applications}}
  \bibinfo{volume}{42}, \bibinfo{number}{6} (\bibinfo{year}{2015}),
  \bibinfo{pages}{3234--3241}.
\newblock


\bibitem[\protect\citeauthoryear{Schlichtkrull, Kipf, Bloem, Van Den~Berg,
  Titov, and Welling}{Schlichtkrull et~al\mbox{.}}{2018}]%
        {schlichtkrull2018modeling}
\bibfield{author}{\bibinfo{person}{Michael Schlichtkrull},
  \bibinfo{person}{Thomas~N Kipf}, \bibinfo{person}{Peter Bloem},
  \bibinfo{person}{Rianne Van Den~Berg}, \bibinfo{person}{Ivan Titov}, {and}
  \bibinfo{person}{Max Welling}.} \bibinfo{year}{2018}\natexlab{}.
\newblock \showarticletitle{Modeling relational data with graph convolutional
  networks}. In \bibinfo{booktitle}{\emph{European Semantic Web Conference}}.
  Springer, \bibinfo{pages}{593--607}.
\newblock


\bibitem[\protect\citeauthoryear{Sharpe}{Sharpe}{1966}]%
        {sharpe1966mutual}
\bibfield{author}{\bibinfo{person}{William~F Sharpe}.}
  \bibinfo{year}{1966}\natexlab{}.
\newblock \showarticletitle{Mutual fund performance}.
\newblock \bibinfo{journal}{\emph{The Journal of business}}
  \bibinfo{volume}{39}, \bibinfo{number}{1} (\bibinfo{year}{1966}),
  \bibinfo{pages}{119--138}.
\newblock


\bibitem[\protect\citeauthoryear{Si, Mukherjee, Liu, Li, Li, and Deng}{Si
  et~al\mbox{.}}{2013}]%
        {si2013exploiting}
\bibfield{author}{\bibinfo{person}{Jianfeng Si}, \bibinfo{person}{Arjun
  Mukherjee}, \bibinfo{person}{Bing Liu}, \bibinfo{person}{Qing Li},
  \bibinfo{person}{Huayi Li}, {and} \bibinfo{person}{Xiaotie Deng}.}
  \bibinfo{year}{2013}\natexlab{}.
\newblock \showarticletitle{Exploiting topic based twitter sentiment for stock
  prediction}. In \bibinfo{booktitle}{\emph{Proceedings of the 51st Annual
  Meeting of the Association for Computational Linguistics (Volume 2: Short
  Papers)}}. \bibinfo{pages}{24--29}.
\newblock


\bibitem[\protect\citeauthoryear{Sutkatti and TORSE}{Sutkatti and
  TORSE}{2019}]%
        {sutkatti2019stock}
\bibfield{author}{\bibinfo{person}{Rashmi Sutkatti} {and} \bibinfo{person}{D
  TORSE}.} \bibinfo{year}{2019}\natexlab{}.
\newblock \showarticletitle{Stock market forecasting techniques: A survey}.
\newblock \bibinfo{journal}{\emph{International Research Journal of Engineering
  and Technology}} \bibinfo{volume}{6}, \bibinfo{number}{5}
  (\bibinfo{year}{2019}).
\newblock


\bibitem[\protect\citeauthoryear{Ticknor}{Ticknor}{2013}]%
        {ticknor2013bayesian}
\bibfield{author}{\bibinfo{person}{Jonathan~L Ticknor}.}
  \bibinfo{year}{2013}\natexlab{}.
\newblock \showarticletitle{A Bayesian regularized artificial neural network
  for stock market forecasting}.
\newblock \bibinfo{journal}{\emph{Expert Systems with Applications}}
  \bibinfo{volume}{40}, \bibinfo{number}{14} (\bibinfo{year}{2013}),
  \bibinfo{pages}{5501--5506}.
\newblock


\bibitem[\protect\citeauthoryear{Vargas, De~Lima, and Evsukoff}{Vargas
  et~al\mbox{.}}{2017}]%
        {vargas2017deep}
\bibfield{author}{\bibinfo{person}{Manuel~R Vargas},
  \bibinfo{person}{Beatriz~SLP De~Lima}, {and} \bibinfo{person}{Alexandre~G
  Evsukoff}.} \bibinfo{year}{2017}\natexlab{}.
\newblock \showarticletitle{Deep learning for stock market prediction from
  financial news articles}. In \bibinfo{booktitle}{\emph{2017 IEEE
  International Conference on Computational Intelligence and Virtual
  Environments for Measurement Systems and Applications (CIVEMSA)}}. IEEE,
  \bibinfo{pages}{60--65}.
\newblock


\bibitem[\protect\citeauthoryear{Vaswani, Shazeer, Parmar, Uszkoreit, Jones,
  Gomez, Kaiser, and Polosukhin}{Vaswani et~al\mbox{.}}{2017}]%
        {vaswani2017attention}
\bibfield{author}{\bibinfo{person}{Ashish Vaswani}, \bibinfo{person}{Noam
  Shazeer}, \bibinfo{person}{Niki Parmar}, \bibinfo{person}{Jakob Uszkoreit},
  \bibinfo{person}{Llion Jones}, \bibinfo{person}{Aidan~N Gomez},
  \bibinfo{person}{{\L}ukasz Kaiser}, {and} \bibinfo{person}{Illia
  Polosukhin}.} \bibinfo{year}{2017}\natexlab{}.
\newblock \showarticletitle{Attention is all you need}. In
  \bibinfo{booktitle}{\emph{Advances in Neural Information Processing
  Systems}}. \bibinfo{pages}{5998--6008}.
\newblock


\bibitem[\protect\citeauthoryear{Veli{\v{c}}kovi{\'{c}}, Cucurull, Casanova,
  Romero, Li{\`{o}}, and Bengio}{Veli{\v{c}}kovi{\'{c}} et~al\mbox{.}}{2018}]%
        {velickovic2018graph}
\bibfield{author}{\bibinfo{person}{Petar Veli{\v{c}}kovi{\'{c}}},
  \bibinfo{person}{Guillem Cucurull}, \bibinfo{person}{Arantxa Casanova},
  \bibinfo{person}{Adriana Romero}, \bibinfo{person}{Pietro Li{\`{o}}}, {and}
  \bibinfo{person}{Yoshua Bengio}.} \bibinfo{year}{2018}\natexlab{}.
\newblock \showarticletitle{{Graph Attention Networks}}.
\newblock \bibinfo{journal}{\emph{International Conference on Learning
  Representations}} (\bibinfo{year}{2018}).
\newblock
\urldef\tempurl%
\url{https://openreview.net/forum?id=rJXMpikCZ}
\showURL{%
\tempurl}
\newblock
\shownote{accepted as poster.}


\bibitem[\protect\citeauthoryear{Wu, Zhang, Shen, and Wang}{Wu
  et~al\mbox{.}}{2018}]%
        {DBLP:conf/cikm/WuZSW18}
\bibfield{author}{\bibinfo{person}{Huizhe Wu}, \bibinfo{person}{Wei Zhang},
  \bibinfo{person}{Weiwei Shen}, {and} \bibinfo{person}{Jun Wang}.}
  \bibinfo{year}{2018}\natexlab{}.
\newblock \showarticletitle{Hybrid Deep Sequential Modeling for Social
  Text-Driven Stock Prediction}. In \bibinfo{booktitle}{\emph{Proceedings of
  the 27th {ACM} International Conference on Information and Knowledge
  Management, {CIKM} 2018, Torino, Italy, October 22-26, 2018}}.
  \bibinfo{pages}{1627--1630}.
\newblock
\urldef\tempurl%
\url{https://doi.org/10.1145/3269206.3269290}
\showDOI{\tempurl}


\bibitem[\protect\citeauthoryear{Xu, Li, Tian, Sonobe, Kawarabayashi, and
  Jegelka}{Xu et~al\mbox{.}}{2018}]%
        {pmlr-v80-xu18c}
\bibfield{author}{\bibinfo{person}{Keyulu Xu}, \bibinfo{person}{Chengtao Li},
  \bibinfo{person}{Yonglong Tian}, \bibinfo{person}{Tomohiro Sonobe},
  \bibinfo{person}{Ken-ichi Kawarabayashi}, {and} \bibinfo{person}{Stefanie
  Jegelka}.} \bibinfo{year}{2018}\natexlab{}.
\newblock \showarticletitle{Representation Learning on Graphs with Jumping
  Knowledge Networks}. In \bibinfo{booktitle}{\emph{Proceedings of the 35th
  International Conference on Machine Learning}}
  \emph{(\bibinfo{series}{Proceedings of Machine Learning Research})},
  Vol.~\bibinfo{volume}{80}. \bibinfo{publisher}{PMLR},
  \bibinfo{address}{Stockholmsmässan, Stockholm Sweden},
  \bibinfo{pages}{5453--5462}.
\newblock


\bibitem[\protect\citeauthoryear{Xu and Cohen}{Xu and Cohen}{2018}]%
        {xu2018stock}
\bibfield{author}{\bibinfo{person}{Yumo Xu} {and} \bibinfo{person}{Shay~B
  Cohen}.} \bibinfo{year}{2018}\natexlab{}.
\newblock \showarticletitle{Stock movement prediction from tweets and
  historical prices}. In \bibinfo{booktitle}{\emph{Proceedings of the 56th
  Annual Meeting of the Association for Computational Linguistics (Volume 1:
  Long Papers)}}, Vol.~\bibinfo{volume}{1}. \bibinfo{pages}{1970--1979}.
\newblock


\bibitem[\protect\citeauthoryear{Yang, Ng, Smyth, and Dong}{Yang
  et~al\mbox{.}}{2020}]%
        {yang2020html}
\bibfield{author}{\bibinfo{person}{Linyi Yang}, \bibinfo{person}{Tin Lok~James
  Ng}, \bibinfo{person}{Barry Smyth}, {and} \bibinfo{person}{Riuhai Dong}.}
  \bibinfo{year}{2020}\natexlab{}.
\newblock \showarticletitle{HTML: Hierarchical Transformer-based Multi-task
  Learning for Volatility Prediction}. In \bibinfo{booktitle}{\emph{Proceedings
  of The Web Conference 2020}}. \bibinfo{pages}{441--451}.
\newblock


\bibitem[\protect\citeauthoryear{Yang, Mo, and Liu}{Yang et~al\mbox{.}}{2015}]%
        {yang2015twitter}
\bibfield{author}{\bibinfo{person}{Steve~Y Yang}, \bibinfo{person}{Sheung
  Yin~Kevin Mo}, {and} \bibinfo{person}{Anqi Liu}.}
  \bibinfo{year}{2015}\natexlab{}.
\newblock \showarticletitle{Twitter financial community sentiment and its
  predictive relationship to stock market movement}.
\newblock \bibinfo{journal}{\emph{Quantitative Finance}} \bibinfo{volume}{15},
  \bibinfo{number}{10} (\bibinfo{year}{2015}), \bibinfo{pages}{1637--1656}.
\newblock


\bibitem[\protect\citeauthoryear{Zhang, Aggarwal, and Qi}{Zhang
  et~al\mbox{.}}{2017}]%
        {zhang2017stock}
\bibfield{author}{\bibinfo{person}{Liheng Zhang}, \bibinfo{person}{Charu
  Aggarwal}, {and} \bibinfo{person}{Guo-Jun Qi}.}
  \bibinfo{year}{2017}\natexlab{}.
\newblock \showarticletitle{Stock price prediction via discovering
  multi-frequency trading patterns}. In \bibinfo{booktitle}{\emph{Proceedings
  of the 23rd ACM SIGKDD International Conference on Knowledge Discovery and
  Data Mining}}. ACM, \bibinfo{pages}{2141--2149}.
\newblock


\bibitem[\protect\citeauthoryear{Zhou, Zhao, and Xu}{Zhou
  et~al\mbox{.}}{2016}]%
        {zhou2016can}
\bibfield{author}{\bibinfo{person}{Zhenkun Zhou}, \bibinfo{person}{Jichang
  Zhao}, {and} \bibinfo{person}{Ke Xu}.} \bibinfo{year}{2016}\natexlab{}.
\newblock \showarticletitle{Can online emotions predict the stock market in
  China?}. In \bibinfo{booktitle}{\emph{International conference on web
  information systems engineering}}. Springer, \bibinfo{pages}{328--342}.
\newblock


\bibitem[\protect\citeauthoryear{Zimbra, Chen, and Lusch}{Zimbra
  et~al\mbox{.}}{2015}]%
        {zimbra2015stakeholder}
\bibfield{author}{\bibinfo{person}{David Zimbra}, \bibinfo{person}{Hsinchun
  Chen}, {and} \bibinfo{person}{Robert~F Lusch}.}
  \bibinfo{year}{2015}\natexlab{}.
\newblock \showarticletitle{Stakeholder analyses of firm-related web forums:
  Applications in stock return prediction}.
\newblock \bibinfo{journal}{\emph{ACM Transactions on Management Information
  Systems (TMIS)}} \bibinfo{volume}{6}, \bibinfo{number}{1}
  (\bibinfo{year}{2015}), \bibinfo{pages}{1--38}.
\newblock


\end{thebibliography}
